\documentclass[amsmath,amssymb,nofootinbib,superscriptaddress,showkeys,twocolumn]{revtex4-2} 
\usepackage{graphicx} 
\usepackage{dcolumn} 
\usepackage{bm} 
\usepackage{subfigure} 
 \usepackage{multirow} 
\usepackage[colorlinks=true, citecolor=blue, urlcolor = magenta, linkcolor= red, bookmarks=true]{hyperref} 
\usepackage{orcidlink} 
\usepackage{booktabs} 

\graphicspath{{figs/}} 

\begin{document}

\title{Probing Gravitational Wave Signatures from Periodic Orbits of Regular Black Holes in Asymptotically Safe Gravity}

\author{Arun Kumar\texorpdfstring{\href{https://orcid.org/0000-0001-8461-5368}{\orcidlink{0000-0001-8461-5368}}{}}}\email{arunbidhan@gmail.com}
\affiliation{Institute for Theoretical Physics and Cosmology, Zhejiang University of Technology, Hangzhou 310023, China}
\author{Abolhassan Mohammadi \texorpdfstring{\href{https://orcid.org/0000-0003-1228-9107}{\orcidlink{0000-0003-1228-9107}}{}}}\email{abolhassanm@gmail.com}
\affiliation{School of Science, Hunan Institute of Technology, Hengyang 421002, China.}
\author{Sushant G. Ghosh \texorpdfstring{\href{https://orcid.org/0000-0002-0835-3690}{\orcidlink{0000-0002-0835-3690}}{}}}\email{sghosh2@jmi.ac.in}
\affiliation{Centre for Theoretical Physics, 
	Jamia Millia Islamia, New Delhi 110025, India}
\affiliation{Astrophysics and Cosmology Research Unit, 
	School of Mathematics, Statistics and Computer Science, University of KwaZulu-Natal, Private Bag 54001, Durban 4000, South Africa}

\begin{abstract}
We investigate bound and periodic timelike geodesics and their associated gravitational wave (GW) signatures in the spacetime of a regular black hole arising in asymptotically safe gravity (ASG). The geometry incorporates quantum corrections via a running gravitational coupling, encoded in a dimensional scaling parameter $\xi$, that modifies the near-horizon structure while preserving asymptotic flatness. Periodic orbits of a stellar-mass compact object around a supermassive black hole are classified using the zoom-whirl taxonomy, and the corresponding GW polarizations are computed within the quadrupole approximation for extreme mass-ratio inspirals (EMRIs). We find that increasing $\xi$ enhances the waveform amplitude and introduces a phase shift relative to the Schwarzschild case, with the peak characteristic strain increasing monotonically with $\xi$. The resulting signals peak in the millihertz band, falling within the sensitivity windows of future space-based detectors, indicating a qualitative path to future parameter constraints on quantum-gravity-induced deviations from classical general relativity in the strong-field regime.

\end{abstract}
	
	
\maketitle

\section{Introduction}\label{sec1}
Black holes, as a prediction of general relativity (GR), are regions of spacetime with an extreme gravitational field in which even light cannot escape. The theory describes black holes as objects with event horizons and singularities, points at which the curvature becomes infinite, leading to the breakdown of the theory. Although black holes are invisible, they are detected by their effect on the surrounding matter. The existence of black holes has been confirmed through multiple observations. The LIGO-Virgo-KAGRA collaboration has detected gravitational waves (GW) from over 100 binary black hole mergers since 2015 \cite{LIGOScientific:2016aoc, LIGOScientific:2016vbw, LIGOScientific:2016vlm, LIGOScientific:2016emj}. Moreover, the first images of a supermassive black hole shadow, in the center of M87 and SgrA*, have been captured by the Event Horizon Telescope \cite{EventHorizonTelescope:2019dse,EventHorizonTelescope:2022wkp}. The observation has opened a new window to our investigation and exploration of the universe. These observations not only confirm black hole existence but also provide unprecedented opportunities to test gravitational theories in the strong-field regime \cite{Berti:2015itd, Yunes:2013dva}.

One of the main sources of GWs is stellar mass black holes (or neutron stars), which move in close orbits around supermassive black holes. They generate extreme mass ratio inspirals (EMRIs), which are the key sources for the future space-based GW detectors, including LISA \cite{Danzmann:1997hm, Schutz:1999xj, Gair:2004iv, LISA:2017pwj, Maselli:2021men}, Taiji \cite{Hu:2017mde}, and TianQin \cite{TianQin:2015yph, Gong:2021gvw}. It is believed that only a small portion of the total energy could be removed by the smaller object through its motion. Therefore, the smaller object can spiral around the supermassive black hole for several years, and this orbital motion can be described by a sequence of periodic orbits (a bound trajectory so that the particle returns to its initial position after an integer number of radial and angular oscillations). These periodic orbits serve as the fundamental building blocks of generic bound trajectories and encode detailed information about the underlying spacetime geometry \cite{Levin:2008ci,Levin:2008mq,Levin:2009sk}. To better understand the dynamics of black hole mergers, it is helpful to classify periodic orbits. All the closed trajectories are categorized by three topological integers as  zoom number ($z$), whirl number ($w$), and vertex number
($v$) \cite{Levin:2008ci,Levin:2008mq,Levin:2009sk}. This classification relies on the ratio of the angular to radial frequencies. This approach has been proven to be a powerful method for categorizing the orbital dynamics and analyzing the waveform from EMRIs, so it has been applied to various spacetimes, such as Schwarzschild and Kerr black holes \cite{Levin:2008ci, Levin:2009sk, Bambhaniya:2020zno, Rana:2019bsn}, charged black holes \cite{Misra:2010pu}, naked singularities \cite{Babar:2017gsg}, Kerr-Sen \cite{Liu:2018vea}, hairy black holes in Horndeski theory \cite{Lin:2023rmo}, and many other black hole spacetimes \cite{Yao:2023ziq, Lin:2022llz, Chan:2025ocy, Wang:2022tfo, Lin:2023eyd, Haroon:2025rzx, Habibina:2022ztd, Zhang:2022psr, Lin:2022wda, Gao:2021arw, Lin:2021noq, Deng:2020yfm, Tu:2023xab, Zhou:2020zys, Gao:2020wjz, Deng:2020hxw, Azreg-Ainou:2020bfl, Wei:2019zdf, Pugliese:2013xfa, Zhang:2022zox, Healy:2009zm, Wang:2025wob, Alloqulov:2025bxh, Wei:2025qlh}, to study how the periodic orbits are affected by the geometry of the black hole. Additionally, the GW emission from the periodic orbits has been considered for different black hole spacetimes \cite{Tu:2023xab, Yang:2024lmj, Shabbir:2025kqh, Junior:2024tmi, Zhao:2024exh, Jiang:2024cpe, Yang:2024cnd, Meng:2024cnq, Li:2024tld, QiQi:2024dwc, Haroon:2025rzx, Alloqulov:2025ucf, Wang:2025hla, Lu:2025cxx, Zare:2025aek, Gong:2025mne, Li:2025sfe, Choudhury:2025qsh, Chen:2025aqh, Deng:2025wzz, Li:2025eln, Zahra:2025tdo}.

Despite the observational successes of the classical GR, black holes admit curvature singularities, signalling the breakdown of the theory \cite{Bardeen:1968proceeding}. This motivates exploring singularity-free alternatives \cite{Hayward:2005gi,Bronnikov:2005gm,Burinskii:2001bq,Fan:2016hvf,Ovalle:2023ref,Mazza:2023iwv,Modesto:2010rv,Casadio:2023iqt, Lewandowski:2022zce,Carballo-Rubio:2019fnb}.  Among various approaches to the alternative gravities, asymptotically safe gravity (ASG) presents a compelling framework that implements quantum corrections via the renormalization group, leading to regular black hole solutions with a single quantum parameter that interpolates between known limits~\cite{Bonanno:2000ep,Torres:2017ygl,Eichhorn:2012va,Pawlowski:2023dda,Stashko:2024wuq,Spina:2024npx}. Recently, Bonanno et al. \citep{Bonanno:2023rzk} have generalized the initial idea of Markov and Mukhanov \cite{Markov:1985py}, and proposed a new model of a regular black hole. They have considered an effective Lagrangian for quantum Einstein gravity, including a multiplicative gravity-matter coupling. By determining the form of the coupling term from ASG, the gravitational collapse of the dust fluid is considered, leading to the direct formation of a regular black hole. The exterior metric is static and spherically symmetric, with the lapse function modified by a logarithmic term involving the scale parameter $\xi$. This parameter encodes quantum corrections and has recently been constrained by observations as $0 < \xi < 0.4565 M^2$ \cite{Stashko:2024wuq}. The regular black hole solution has been studied for different topics \cite{Stashko:2024wuq,Spina:2024npx,Bhattacharjee:2025xcb,Mustafa:2025cou,Mannobova:2025uqf,Urmanov:2025nou,Zhao:2025sck,Turakhonov:2025ojy}, in particular, in a recent study, Gao examined the metric for strong lensing and optical appearances \cite{Gao:2024cgg}.

In this work, we perform a detailed investigation of time-like geodesic motion in the regular black hole spacetime arising from gravitational dust collapse in ASG. Starting from the modified exterior metric, we derive the effective potential and determine the innermost stable circular orbits (ISCOs) and marginally bound orbits (MBOs), emphasizing their dependence on the quantum correction parameter. We then classify the periodic orbits using the $(z, w, v)$ taxonomy, and construct them for different sets of zoom, whirl, and vertex numbers. Based on these results, the associated gravitational-wave polarizations, waveform, and characteristic strain are computed. It is found that the quantum correction leaves observable imprints on both the orbital dynamics and the emitted gravitational radiation, leading to an enhancement of the characteristic strain. The resulting characteristic strain crosses the sensitivity curves of LISA, Taiji, and TianQin, providing an opportunity to probe and constrain quantum effects using observational data.


The paper is organized as follows. In Sec.~\ref{sec:metric_model}, we briefly introduce the spacetime metric of the regular black hole in ASG (RBHASG). We then derive the effective potential governing timelike geodesics and analyze the MBOs and ISCOs, highlighting their dependence on the quantum scaling parameter \(\xi\). In Sec.~\ref{sec:periodic_orbits}, we classify and construct periodic orbits using the \((z,w,v)\) taxonomy for different values of zoom, whirl, and vertex numbers. Section~\ref{sec:gw} is devoted to computing the GW signal, including the two polarization modes \(h_{+}\) and \(h_{\times}\), the corresponding waveform via discrete Fourier transform, and the characteristic strain. We also compare the characteristic strain with the sensitivity curves of future space-based detectors LISA, Taiji, and TianQin. Finally, we summarize our main findings and conclusions in Sec.~\ref{sec:conclusion}. Throughout this paper, we use geometric units with \(G = c = 1\) unless otherwise specified.

\section{Regular Black Hole in ASG and Geodesics}\label{sec:metric_model}
\begin{figure}[hb!] 
\centering
\begin{tabular}{c c}
	\includegraphics[width=0.45\textwidth]{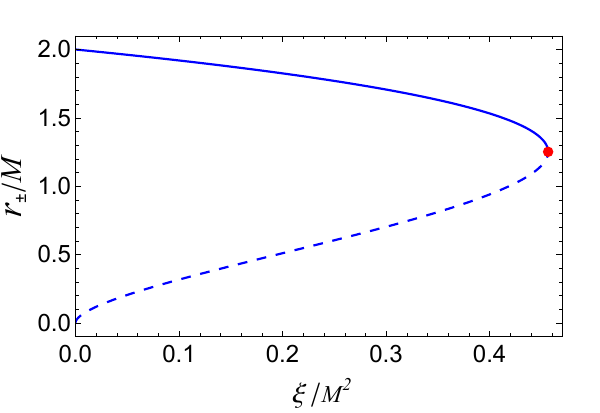}\\
    \end{tabular}
    \caption{The event horizon (solid curve) and the Cauchy horizon (dashed curve) of RBHASG with parameter $\xi$. We have kept $M=1$.}
\label{horizon}
\end{figure}

We begin by briefly introducing the static, spherically symmetric regular black hole metric derived from the gravitational collapse of dust in ASG, as first proposed by \citep{Bonanno:2023rzk}. This metric originates from an effective action that includes a coupling between gravity and matter, which reads 
\begin{equation}
    S = \frac{1}{16\pi G_N} \int d^4x \sqrt{-g} \; \big( \; R + 2 \chi(\epsilon) \mathcal{L} \; \big),
\end{equation}
where $G_N$ is the Newton constant and the coupling between gravity and matter is represented by $\chi(\epsilon)$ in which $\chi(0) = 8\pi G_N$. $\mathcal{L}$ is the matter Lagrangian with proper density $\epsilon$. The resulting field equation is obtained as
\begin{equation}
    R_{\mu\nu} - \frac{1}{2} g_{\mu\nu} R = 8\pi G(\epsilon) T_{\mu\nu} - \Lambda(\epsilon) g_{\mu\nu},
\end{equation}
in which $G(\epsilon)$ and $\Lambda(\epsilon)$ are the effective Newton constant and the cosmological constant, respectively, given by 
\begin{equation}
    G(\epsilon) = \frac{\partial \big(\epsilon \;\chi\big)}{\partial \epsilon}, \quad \Lambda(\epsilon) = -\frac{\partial \chi}{\partial \epsilon} \epsilon^2.
\end{equation}
Assuming that the behavior of $G(\epsilon)$ is governed by a renormalization group trajectory close to the ultraviolet fixed point of the Asymptotic Safe program, the running of $G(\epsilon)$ can be approximated by
\begin{equation}
    G(\epsilon) = \frac{G_N}{1 + \xi \epsilon},
\end{equation}
in which $\xi$ is a scale parameter; refer to  \cite{Bonanno:2019ilz,Bonanno:2021squ,Bonanno:2023rzk} for more detail. Although the exact value of the $\xi$ parameter is not known yet, it can be constrained by observations. In the recent study \cite{Bonanno:2023rzk}, it is determined that the parameter $\xi$ stands between $0<\xi<0.4565 M^2$. For the gravitational collapse of dust, the resulting exterior spherically symmetric metric is obtained as \cite{Bonanno:2023rzk}
\begin{equation}\label{metric}
    ds^2 = -f(r) dt^2 + \frac{1}{f(r)} dr^2 + r^2 d\Omega^2
\end{equation}
where the function $f(r)$ is given by 
\begin{equation}
    f(r) = 1 - \frac{r^2}{3 \xi} \; \log \left( 1 + \frac{6M \xi}{r^3} \right),
\end{equation}
so that $M$ represents the mass of the configuration. 
Next, we check the consistency of the regular black hole in ASG (RBHASG) (\ref{metric}). 
Expanding Eq.~(6) at large radii yields $f(r) \simeq 1 - 2M/r + 6M^2\xi/r^4 + \mathcal{O}(\xi^2/r^7)$, showing that quantum corrections decay rapidly as $\sim r^{-4}$. Consequently, deviations from Schwarzschild remain strongly suppressed in the infrared regime, while near the strong-field region $r \sim \mathcal{O}(M)$ they scale as $\delta f/f \sim \mathcal{O}(\xi/M^2)$, leading to small but controlled modifications of orbital dynamics. 
For large $r$,  we find that the lapse function of the metric (\ref{metric}) reduces to $f(r) \approx 1 - 2M/r$. Thus, in the limit $r \to \infty$, the spacetime approaches the Schwarzschild solution, confirming asymptotic flatness. The quantum parameter $\xi$ therefore introduces corrections primarily in the near-horizon region while preserving the classical limit at large distances. We have shown how the horizon radii (the event horizon $r_+$ and the Cauchy horizon $r_-$) of RBHASG change with parameter $\xi$ in Fig. \ref{horizon}, where solid and dashed curves represent the event and Cauchy horizon, respectively. The red dot represents the extremal black hole where the event horizon ($r_+$) is equal to the Cauchy horizon ($r_-$).
\subsection{Time-Like Geodesics}
The motion of a massive test particle in a static and spherically symmetric spacetime is governed by the geodesic equations, which can be obtained from the Lagrangian formulation of the metric \cite{Chandrasekhar:1985kt, Wald:1984rg, Misner:1973prb}. The corresponding Lagrangian is given by
\begin{eqnarray}\label{Lagrangian}
   2\mathcal{L} & = &- f(r) \;\dot{t}^2 + \frac{\dot{r}^2}{f(r)} + r^2(\dot{\theta}^2+\sin \theta^2\dot{\phi}^2)
\end{eqnarray} 
where the overdot signifies the differentiation with respect to the affine parameter $\tau$, and $r$ and $t$ denote dimensionless coordinates obtained via the rescaling $r\to r/M$ and  $t\to t/M$, respectively. $\xi$ has a dimension of length squared. After rescaling by $M$, we treat $\xi/M^2$ as dimensionless. The normalization condition $2\mathcal{L}=-1$ corresponds to timelike geodesics. For further discussions, we shall confine ourselves to the equatorial plane $\theta=\pi/2$, and hence $\dot{\theta}=0$. Now, by employing the Euler-Lagrange equations to the Lagrangian \eqref{Lagrangian}, we obtain two conserved quantities associated with the Killing vectors generating time translation, $\partial_t$, and axial symmetry, $\partial_\phi$, given by \cite{Chandrasekhar:1985kt}
\begin{equation}\label{conserevQ}
L= r^2\dot{\phi},  \qquad 
E=\left(1-\frac{r^2}{3\xi} \log \left(1 + \frac{6 \xi }{r^3}\right)\right)\dot{t}.
\end{equation}
Substituting the conserved quantities \eqref{conserevQ} in Lagrangian \eqref{Lagrangian} and employment of the normalization condition $2\mathcal{L}=-1$, leads to the following radial equation
\begin{equation}\label{radial}
     \dot{r}^2 = E^2 - f(r) \left( 1 + \frac{L^2}{r^2} \right) 
    \equiv E^2-V_{\rm eff}(r),
\end{equation} 
where $V_{\rm eff}(r)$ represents the effective potential that reads
\begin{equation}\label{Veff}
   V_{\rm eff}(r)=\left(1+\frac{L^2}{r^2}\right)\left(1-\frac{r^2}{3\xi} \log \left(1 + \frac{6 \xi }{r^3}\right)\right)
\end{equation}
From the structure of the effective potential $V_{\rm eff}$, it follows that 
$\lim_{r \to \infty} V_{\rm eff}(r) = 1,$
which confirms that the spacetime is asymptotically flat  \cite{Chandrasekhar:1985kt, Wald:1984rg}.  Additionally, the energy $ E=1$ is treated as the critical value between bounded and unbounded orbits, so particles with energy $E>1$ can escape to infinity, corresponding to unbounded orbits. On the other hand, for particles with energy $E < 1$, we have bounded orbits. The bounded orbits exist between MBOs and ISCOs, so that a massive particle with a bounded orbit should have an energy between $E_{\rm ISCO} \leq E \leq E_{\rm MBO}$. The energy $E_{\rm ISCO}$ stands for the energy of ISCO. It is the lower energy limit for the stable bound orbits, so that particles with energy less than this value can not maintain circular motion and will fall into the black hole. Moreover, the energy $E_{\rm MBO} = 1$ is the energy of MBOs, stating that the particle has exactly the energy needed to escape to infinity if slightly perturbed outward. In addition, the condition for the angular momentum is $L \geq L_{\rm ISCO}$ to maintain a bound orbit. 
\begin{figure}[ht!] 
\centering
\begin{tabular}{c c}
	\includegraphics[width=0.45\textwidth]{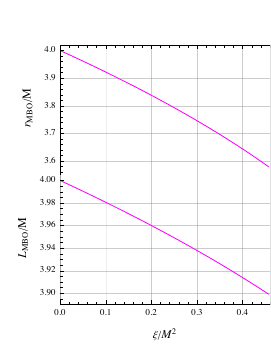}\\

    \end{tabular}
    \caption{Behavior of the MBO parameters as functions of the quantum parameter $\xi$: radius $r_{\mathrm{MBO}}$ (upper panel) and angular momentum $L_{\mathrm{MBO}}$ (lower panel) for the RBHASG.}
\label{mbo}
\end{figure}
\begin{figure}[]
\centering
\begin{tabular}{c c c}
	\includegraphics[width=0.45\textwidth]{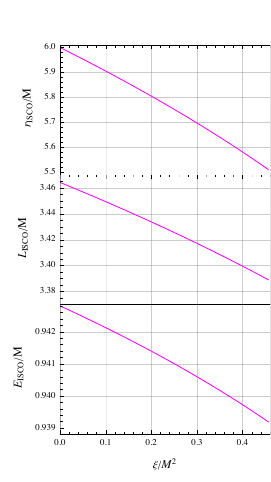}
    \end{tabular}
    \caption{Dependence of ISCO parameters on the quantum parameter $\xi$: radius $r_{\mathrm{ISCO}}$  (upper panel), angular momentum $L_{\mathrm{ISCO}}$ (middle panel), and energy $E_{\mathrm{ISCO}}$ (lower panel) for RBHASG.}
\label{isco}
\end{figure}

\subsection{MBOs and ISCOs}
The MBO corresponds to an unstable circular orbit with specific energy $E=1$, which represents the threshold between bound and unbound motion \cite{Chandrasekhar:1985kt, Bardeen:1972fi, Shapiro:1983du}. It is the circular orbit with the smallest radius for which the particle remains marginally bound. The radius $r_{\rm MBO}$ and the corresponding angular momentum $L_{\rm MBO}$ are determined from the following conditions
\begin{itemize}
    \item [i)] $V_{\rm eff}(r)\Big|_{r= r_{\rm MBO}} = 1$ 
    \item [ii)] $\frac{dV_{\rm eff}(r)}{dr}\Big|_{r= r_{\rm MBO}} = 0$ 
\end{itemize}
Due to the nonlinearity of the equations, $r_{\rm MBO}$ and $L_{\rm MBO}$ are obtained numerically. Figure~\ref{mbo} shows the dependence of $r_{\rm MBO}$ (upper panel) and $L_{\rm MBO}$ (lower panel) on the parameter $\xi$. Both quantities decrease with increasing $\xi$, indicating that the quantum corrections shift the MBOs inward.

The ISCO defines the transition between stable and unstable circular motion and plays a fundamental role in black hole accretion physics \cite{Bardeen:1972fi, Chandrasekhar:1985kt, Wald:1984rg}. For radii smaller than $r_{\rm ISCO}$, circular orbits become unstable and the particle plunges into the black hole. The radius, angular momentum, and energy of the ISCO are obtained by solving the conditions
\begin{itemize}
    \item [i)] $V_{\rm eff}(r)\Big|_{r= r_{\rm ISCO}} = E^2$ 
    \item [ii)] $\frac{dV_{\rm eff}(r)}{dr}\Big|_{r= r_{\rm ISCO}} = 0$ 
    \item [iii)] $\frac{d^2V_{\rm eff}(r)}{dr^2}\Big|_{r= r_{\rm ISCO}} = 0$ 
\end{itemize}
where the first two conditions guarantee circular motion and the third imposes marginal stability \cite{Chandrasekhar:1985kt}. These equations are solved numerically. Figure~\ref{isco} presents the ISCO radius $r_{\rm ISCO}$, angular momentum $L_{\rm ISCO}$, and energy $E_{\rm ISCO}$ as functions of $\xi$. We observe that all three quantities decrease as $\xi$ increases, demonstrating that the quantum parameter reduces the size and energy scale of the ISCOs.

\begin{figure}[ht!] 
\centering
\begin{tabular}{c c}
	\includegraphics[width=0.45\textwidth]{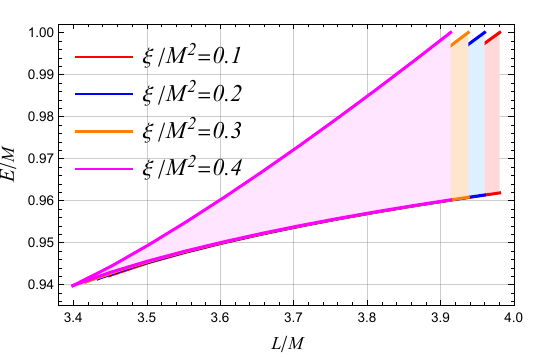}\\
    \end{tabular}
    \caption{Allowed parameter space $(E, L)$ for bound timelike orbits around RBHASG for different values of the quantum parameter $\xi$. The shaded regions indicate where stable bound orbits exist, bounded by the ISCO (lower boundary) and MBO (upper boundary).}
\label{EL}
\end{figure}

Since stable bound circular orbits exist only for angular momentum values between those corresponding to the ISCO and the MBO \cite{Chandrasekhar:1985kt, Bardeen:1972fi, Wald:1984rg}, we parametrize the specific angular momentum as
\begin{equation}\label{Momentum}
  L=L_{\rm ISCO}+\epsilon(L_{\rm MBO}-L_{\rm ISCO}), 
\end{equation} 
where the dimensionless parameter $\epsilon$ satisfies $0\leq\epsilon\leq1$. The limiting cases $\epsilon=0$ and $\epsilon=1$ correspond to the ISCO and MBO, respectively, ensuring that the analysis is restricted to bound orbits. Using this parametrization, we construct the allowed $E$–$L$ domain for bound orbits around the  RBHASG. Figure~\ref{EL}  depicts the $E$–$L$ regions for different scaling parameter $\xi$. We observe that the specific energy $E$ increases monotonically with increasing angular momentum $L$, consistent with the general behavior of circular geodesics in spherically symmetric spacetimes \cite{Chandrasekhar:1985kt}. Furthermore, the allowed $E$–$L$ region broadens as $\xi$ increases, indicating that quantum corrections enlarge the parameter space for bound orbital motion.

\section{Periodic Orbits}\label{sec:periodic_orbits}
In this section, we investigate the periodic orbits around the RBHASG. To construct the particle trajectories, one may in principle integrate Eqs.~\eqref{conserevQ} and \eqref{radial}, which determine $t(\tau)$, $\phi(\tau)$, and $r(\tau)$ in terms of the affine parameter $\tau$ \cite{Chandrasekhar:1985kt, Levin:2008mq}. However, the radial equation \eqref{radial} contains a square root, thereby requiring an explicit choice of sign corresponding to inward ($\dot r<0$) and outward ($\dot r>0$) motion. To avoid this ambiguity, we obtained a convenient equation of motion by differentiating Eq. \eqref{radial}, which reads
\begin{equation}\label{radial1}
    \ddot{r}=\frac{L^2f(r)}{r^3}-\frac{f'(r)V_{\rm eff}(r)}{2f(r)}.
\end{equation} 
This radial equation is convenient for numerical integration, as it avoids sign ambiguities and directly describes the dynamical evolution of the orbit. Moreover, it provides a transparent framework for analyzing the stability of circular orbits and their transition to more intricate periodic or zoom--whirl trajectories in the strong-field regime \cite{Levin:2008mq, Glampedakis:2002cb}. 

\begin{figure}[ht!] 
\centering
\begin{tabular}{c c}
	\includegraphics[width=0.45\textwidth]{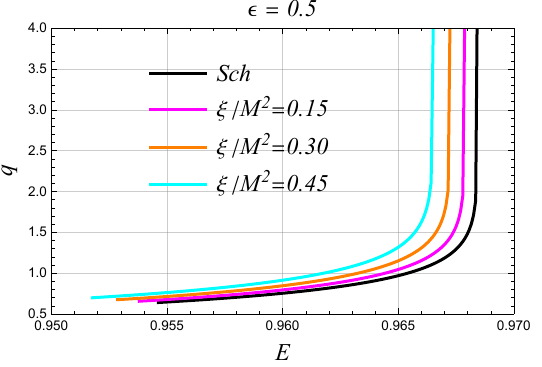}\\
    \includegraphics[width=0.45\textwidth]{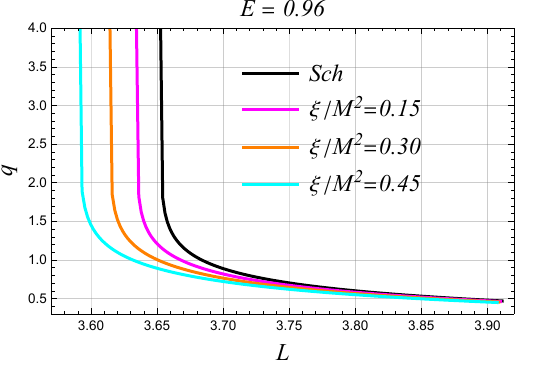}
    \end{tabular}
    \caption{\textcolor{blue}{Behavior} of the rational frequency ratio $q = \omega_\phi/\omega_r - 1$ as a function of energy $E$ for fixed $\epsilon = 0.5$ (upper panel) and angular momentum $L$ for fixed $E = 0.96$ (lower panel), shown for different values of $\xi$.}
\label{qLE}
\end{figure}
To classify the different periodic orbits, Levin and Perez--Giz \cite{Levin:2008mq} introduced an elegant taxonomy in which each periodic orbit is uniquely identified by a triplet of integers $(z,w,v)$. These integers represent the number of zooms ($z$), whirls ($w$), and vertices ($v$) traced by the particle during one radial period. The frequency ratio $q$ defined through the azimuthal ($\omega_{\phi}$) and radial ($\omega_{r}$) oscillation frequencies, is expressed as
\begin{equation}\label{qequation}
    q=\frac{\omega_{\phi}}{\omega_{r}}-1 \equiv w+\frac{v}{z}.
\end{equation}
The parameter $q$ measures the accumulated periapsis advance beyond that of a simple closed ellipse and therefore encodes the geometric structure of the orbit in the strong-field regime \cite{Levin:2008mq, Glampedakis:2002cb}. 
The periodicity of the orbits arises as the ratio of the azimuthal to radial frequencies, $\omega_{\phi}/\omega_{r}$, becomes a rational number. In general, orbital motion is characterized by an irrational frequency ratio; however, it can always be approximated to arbitrary accuracy by rational numbers. The frequency ratio is linked to the equatorial angle via
\begin{equation}\label{eqAngle}
    \frac{\omega_{\phi}}{\omega_{r}}=\frac{\Delta \phi}{2\pi},
\end{equation} 
with the quantity $\Delta \phi=2\oint d\phi\equiv 2\int_{r_1}^{r_2}\dot{\phi}/{\dot{r}}$ represents the total equatorial angle accumulated during a single radial cycle and is required to be an integer multiple of $2\pi$. By using the above relations alongside the equations of motion, we can also calculate $q$ in terms of the spacetime parameters as 
\begin{eqnarray}\label{qequation1}
    q&=&\frac{1}{\pi}\int_{r_1}^{r_2}\frac{L}{r^2\sqrt{E^2-V_{\rm eff}(r)}}-1
\end{eqnarray} 
with $r_1$ and $r_2$ being the periapsis and apoapsis radii of the periodic orbits, respectively. 

The periapsis $r_1$ and apoapsis $r_2$ are determined from the positive real roots of the equation $E^2-V_{\rm eff}(r)=0$. For a bound orbit, both turning points must exist and be real. Since this condition is not satisfied for arbitrary choices of the energy $E$ and angular momentum $L$, only those values of $E$ (for a fixed $L$) or $L$ (for a fixed $E$) that yield real turning points are considered. The corresponding orbital integrals are then evaluated over this physically allowed parameter range.

\begin{figure*}[ht!] 
\centering
    \includegraphics[width=0.32\textwidth]{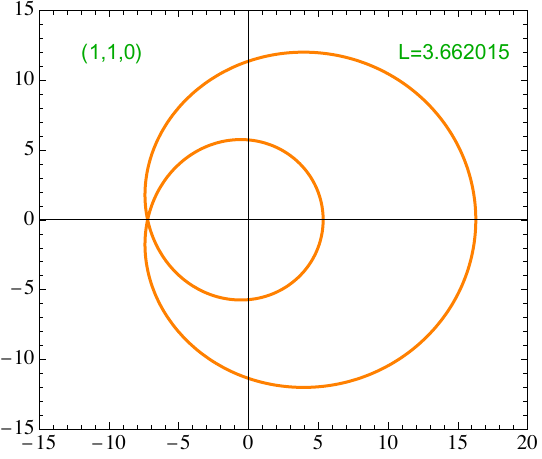}
    \includegraphics[width=0.32\textwidth]{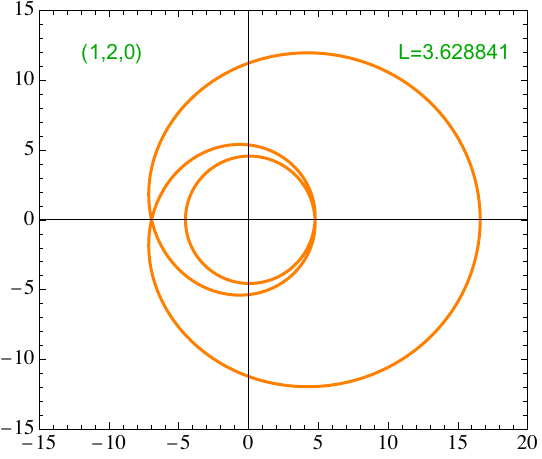}
    \includegraphics[width=0.32\textwidth]{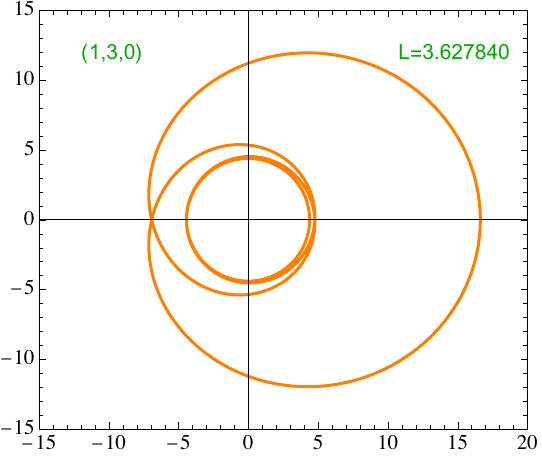}\\

    \vspace{0.2cm}
    \includegraphics[width=0.32\textwidth]{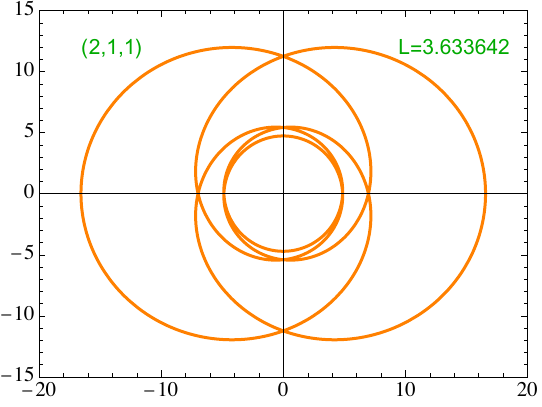}
    \includegraphics[width=0.32\textwidth]{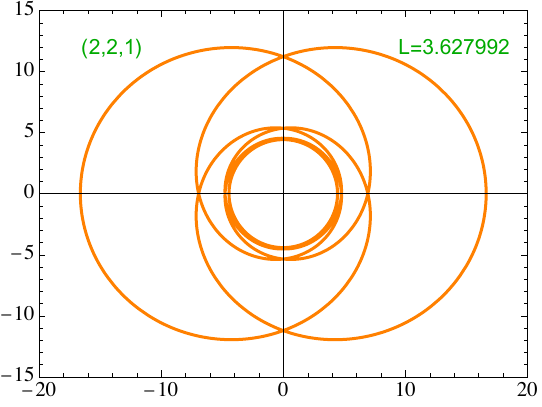}
    \includegraphics[width=0.32\textwidth]{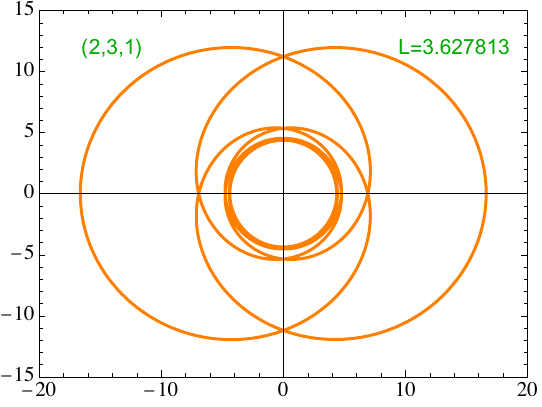}\\

    \vspace{0.2cm}
    \includegraphics[width=0.32\textwidth]{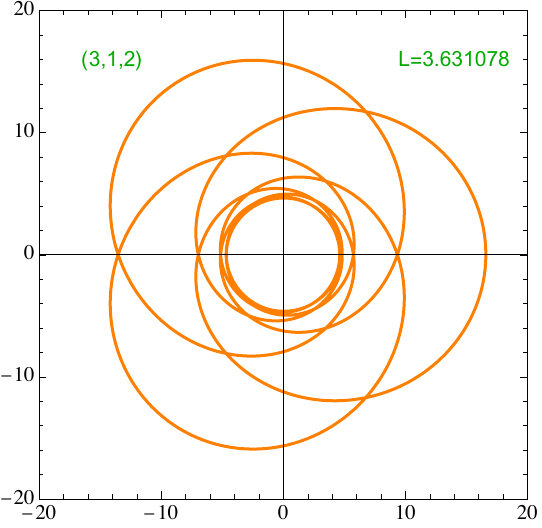}
    \includegraphics[width=0.32\textwidth]{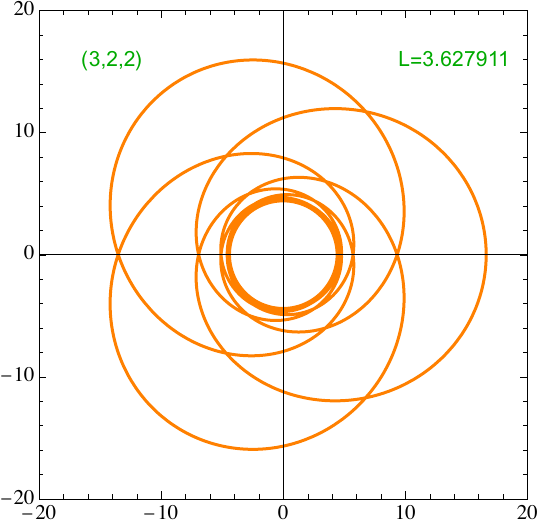}
    \includegraphics[width=0.32\textwidth]{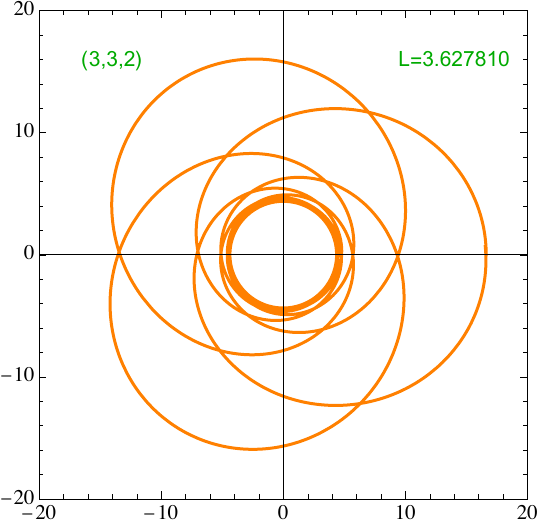}\\

    \vspace{0.2cm}
    \includegraphics[width=0.32\textwidth]{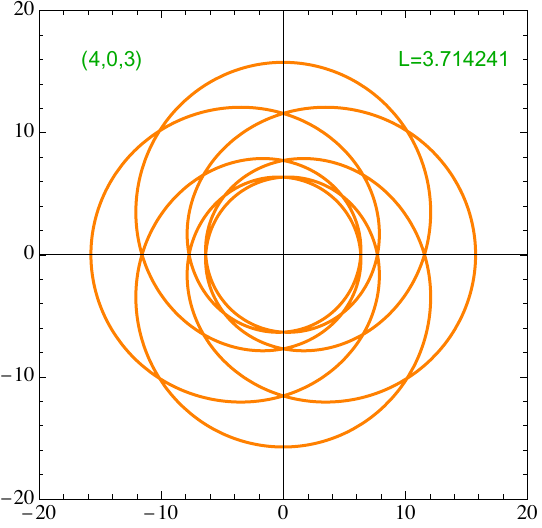}
    \includegraphics[width=0.32\textwidth]{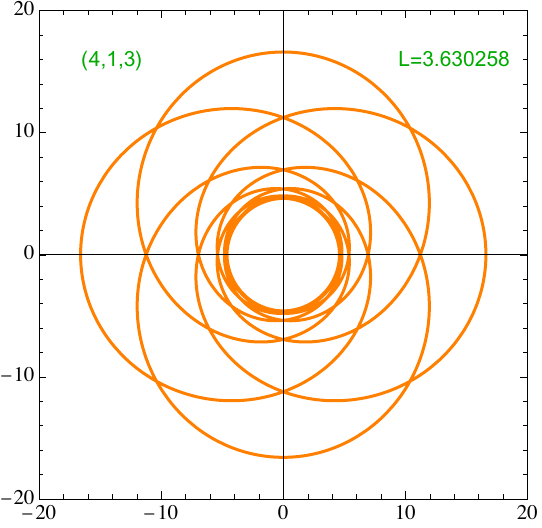}
    \includegraphics[width=0.32\textwidth]{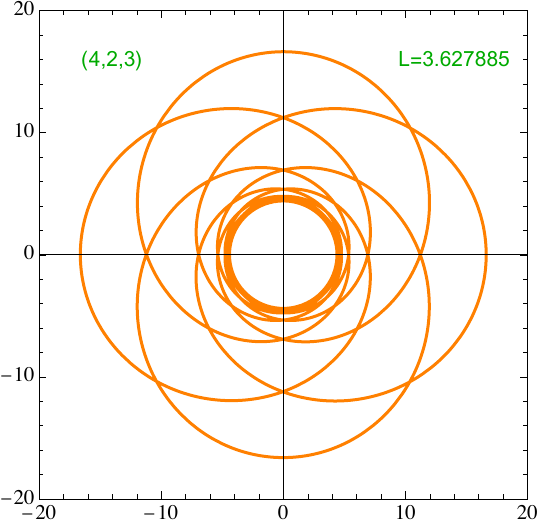}
    \caption{Periodic orbits around RBHASG with $\xi = 0.2M^2$ and fixed energy $E = 0.96$, classified according to their topological integers $(z, w, v)$. Each panel shows trajectories for different zoom ($z$), whirl ($w$), and vertex ($v$) numbers, with the corresponding angular momentum $L$ indicated. Increasing $z$ produces more complex orbital structures, while increasing $w$ results in additional loops around the black hole before reaching the apoapsis.}
\label{or1}
\end{figure*}

The dependence of the frequency parameter $q$ on the specific energy $E$ (for fixed $\epsilon=0.5$) and on the angular momentum $L$ (for fixed $E=0.96$) is shown in Fig.~\ref{qLE} for several values of the parameter $\xi$. As illustrated in the upper panel in Fig.~\ref{qLE}, $q$ increases monotonically with the energy $E$, reflecting the enhancement of relativistic periapsis precession as the orbit approaches the separatrix in the strong-field regime \cite{Levin:2008mq, Glampedakis:2002cb}. The growth of $q$ eventually saturates as $E$ approaches its maximum allowed value for bound motion. This maximum energy exhibits an inverse dependence on $\xi$, decreasing as $\xi$ increases, indicating that quantum corrections reduce the available energy window for highly precessing bound orbits.
In contrast, the lower panel of Fig.~\ref{qLE} shows that $q$ decreases monotonically with increasing angular momentum $L$. Larger angular momentum shifts the orbit outward, weakening relativistic precession effects and thus reducing the frequency ratio $q$, until the minimum allowed value of $L$ is reached at the ISCO \cite{Chandrasekhar:1985kt}. 
Representative periodic orbits corresponding to different combinations of zoom ($z$), whirl ($w$), and vertex ($v$) numbers are displayed in Fig.~\ref{or1}. As expected from the Levin--Perez-Giz classification \cite{Levin:2008mq}, the orbital structure becomes increasingly intricate with growing zoom number $z$, while a larger whirl number $w$ leads to additional near-circular loops around the black hole during each radial cycle. This progressive complexity reflects the strong-field dynamics characteristic of relativistic bound motion near the innermost stable orbit.
 \begin{table*}[htb!]
 {\footnotesize
\resizebox{\textwidth}{!}{ 
 \begin{centering}	
	\begin{tabular}{|c c c c c c c c c c c c c c|}

\hline
{$\mathcal{\xi}$ } & {$L$}& {$E_{(1,1,0)}$}& {$E_{(1,2,0)}$} &{$E_{(1,3,0)}$} & {$E_{(2,1,1)}$} & {$E_{(2,2,1)}$} &{$E_{(2,3,1)}$} & {$E_{(3,1,2)}$ }& {$E_{(3,2,2)}$} &{$E_{(3,3,2)}$} &{$E_{(4,0,3)}$ } &{$E_{(4,1,3)}$ }&{$E_{(4,2,3)}$} \\ \hline
$0$&$3.7320$& $0.9654$ &$0.9683$&$0.9684$& $0.9680$& $0.9684$&$0.9684$ & $0.9682$& $0.9684$&$0.9684$ &$0.9599$ & $0.9682$&$0.9684$\\
\hline
$0.1$&$3.7149$& $0.9647$ &$0.9680$& $0.9680$& $0.9676$& $0.9680$& $0.9680$& $0.9678$ &$0.9680$&$0.9680$& $0.9589$& $0.9678$&$0.9680$ \\

\hline
$0.2$&$3.6968$& $0.9640$ &$0.9676$& $0.9676$& $0.9671$& $0.9676$& $0.9676$ & $0.9673$& $0.9676$ &$0.9676$& $0.9578$ & $0.9674$&$0.9676$\\
\hline
{$0.3$}&$3.6775$& $0.9632$ &$0.9671$& $0.9672$& $0.9666$& $0.9672$&$0.9672$& $0.9669$ &$0.9672$&$0.9672$& $0.9566$& $0.9670$&$0.9672$ \\
\hline
{$0.4$}&$3.6567$& $0.9623$ &$0.9666$& $0.9667$& $0.9660$& $0.9667$& $0.9667$& $0.9663$ &$0.9667$&$0.9667$& $0.9552$& $0.9664$&$0.9667$ \\
       \hline
	\end{tabular} 
\end{centering}}
	\caption{Numerical results of the energy for periodic orbits around RBHASG for different values of the parameter $\xi$. We kept $\epsilon=0.5$ for these results.}\label{table2}
    }
\end{table*}

\section{Gravitational Waves}\label{sec:gw}
The EMRIs are among the primary targets of future space-based GW detectors such as LISA, Taiji, and TianQin \cite{LISA:2017pwj, Hu:2017mde, TianQin:2015yph}. An EMRI system typically consists of a stellar-mass compact object (e.g., a black hole or neutron star with mass $\sim 1$--$100\,{\rm M_\odot}$) orbiting a supermassive black hole of mass $\sim 10^{6}$--$10^{7}\,{\rm M_\odot}$. Because of the large mass hierarchy and prolonged inspiral phase, such systems emit GWs within the millihertz frequency band, which lies well within the sensitivity windows of these detectors.
The emitted gravitational radiation encodes detailed information about the strong-field orbital dynamics and the underlying spacetime geometry of the central black hole \cite{Barack:2003fp,Gair:2012nm}. Consequently, EMRIs offer a unique opportunity to test GR in the strong-gravity regime and to probe possible deviations arising from quantum-gravity-motivated corrections. In particular, within the framework of ASG, the RBHASG solution (\ref{metric}) characterized by the scale parameter $\xi$ modifies the geodesic structure and orbital frequencies. These modifications can, in turn, imprint observable signatures on the form relative to the classical Schwarzschild case. Future high-precision observations may therefore detect or constrain such deviations, placing bounds on the quantum correction parameter $\xi$.
\begin{figure*}[ht!] 
\centering
	\subfigure{\includegraphics[width=0.45\textwidth]{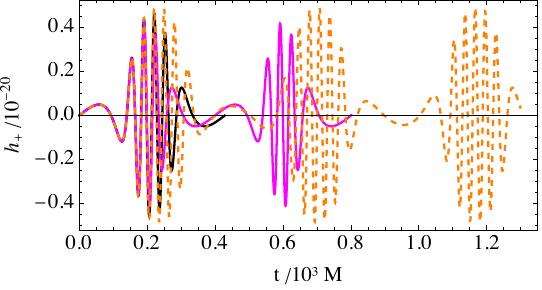}} 
    \subfigure{\includegraphics[width=0.45\textwidth]{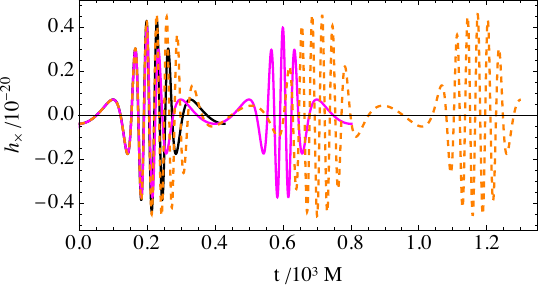}}
    \caption{\textcolor{blue}{Waveforms} (plus polarization $h_+$ in upper panel, cross polarization $h_\times$ in lower panel) for different periodic orbits around RBHASG with $\xi = 0.2M^2$ and $E = 0.96$. The orbits shown are $(1,2,0)$ (black), $(2,1,1)$ (magenta), and $(3,2,2)$ (dashed orange). The system parameters are $m = 10M_{\odot}$ for the stellar-mass object, $M = 10^6M_{\odot}$ for the supermassive black hole, inclination $\iota = \zeta = \pi/4$, and luminosity distance $D_L = 200$ Mpc.}
\label{gw1}
\end{figure*}

\begin{figure*}[ht!]
  \centering
  \begin{tabular}{ c }
    \includegraphics[width=0.4\textwidth]{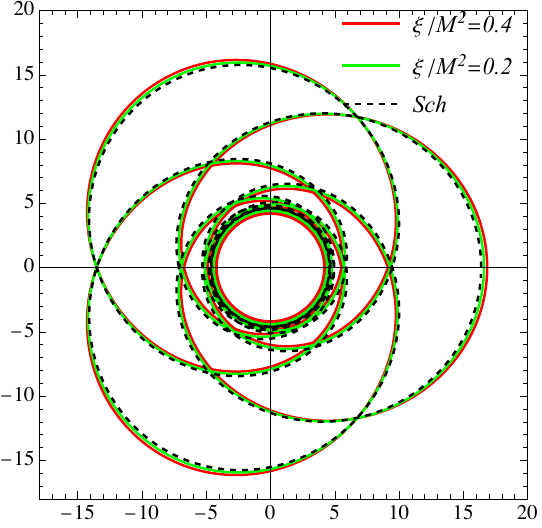}
  \end{tabular}%
  \begin{tabular}{ c c }
   \includegraphics[width=0.5\textwidth]{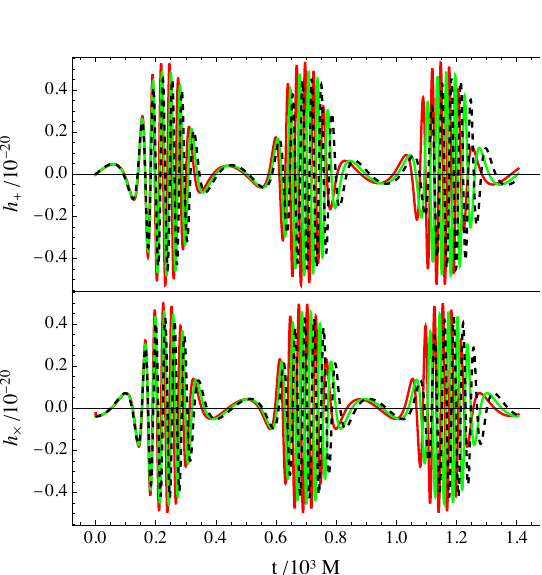}\\ 
     \end{tabular}

\caption{Effect of the quantum parameter $\xi$ on the $(3,2,2)$ periodic orbit (left panel) and its corresponding GW signals $h_+$ and $h_\times$ (right panels) for fixed $E = 0.96$. Results are shown for $\xi = 0$ (Schwarzschild), $\xi = 0.2M^2$, and $\xi = 0.4M^2$. Increasing $\xi$ enhances the waveform amplitude and introduces a discernible phase shift.}
\label{orgw}
\end{figure*}
\begin{figure*}[ht!]
  \centering
  \begin{tabular}{ c }
    \includegraphics[width=0.4\textwidth]{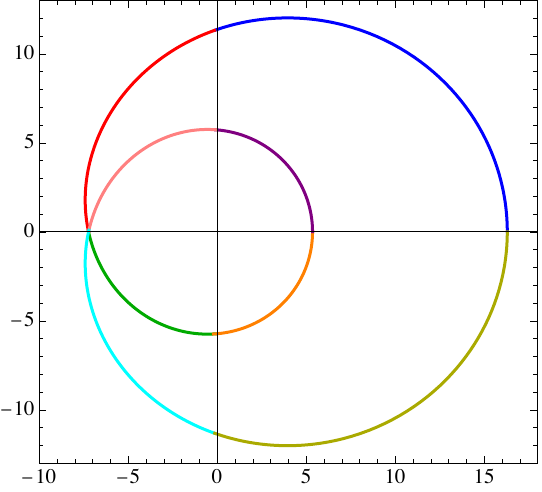}
  \end{tabular}%
  \begin{tabular}{ c c }
   \includegraphics[width=0.45\textwidth]{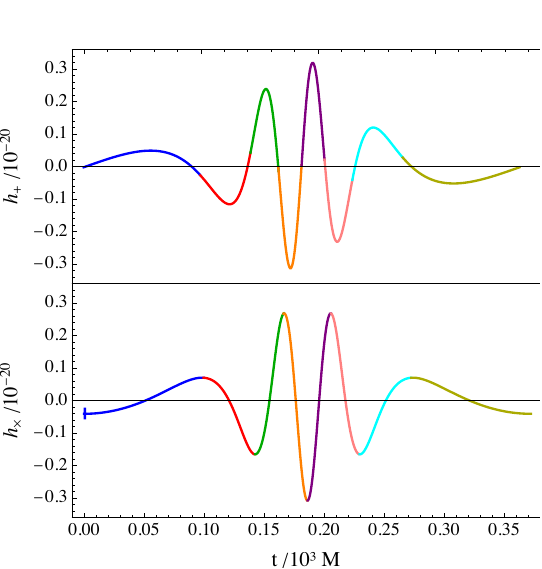}\\ 
     \end{tabular}

\caption{Correlation between orbital segments and GW features for the $(1,1,0)$ periodic orbit around RBHASG with $\xi = 0.2M^2$ and $E = 0.96$. Different colors highlight corresponding portions of the trajectory (left panel) and the resulting plus and cross polarization waveforms (right panels), illustrating how specific orbital phases map to features in the GW signal.}
\label{orgw1}
\end{figure*}
In computing the waveform from EMRIs, we adopt the adiabatic approximation, which is well justified by the system's extreme mass ratio \cite{Mino:1997bx, Poisson:2011nh, Barack:2009ux}. In this regime, the radiation-reaction (inspiral) timescale is much longer than the orbital period, implying that the specific energy and angular momentum of the smaller compact object evolve slowly over multiple orbital cycles. As a result, the motion can be approximated as a sequence of geodesics in the background black hole spacetime. In the present analysis, we neglect the self-force backreaction of gravitational radiation on the trajectory of the smaller object and focus on geodesic motion. This approximation captures the leading-order features of the orbital dynamics and is commonly employed in exploratory studies of strong-field effects \cite{Barack:2009ux, Poisson:2011nh}.  To model the waveform emitted from periodic orbits around the regular black hole, we employ the numerical kludge (NK) approach \cite{Babak:2006uv, Gair:2005ih}. This semi-relativistic framework integrates geodesic trajectories in curved spacetime with flat-spacetime wave-generation formulas, providing an efficient, reliable tool for capturing qualitative strong-field features of EMRI waveforms. The NK method has been widely used in the literature to investigate orbital dynamics and environmental or beyond-GR effects on GW signatures \cite{Babak:2006uv,Gair:2005ih,Alloqulov:2025bxh, Haroon:2025rzx,Ahmed:2025azu}.

It should be noted that the NK framework, by construction, combines geodesic trajectories computed in the curved background spacetime with flat-spacetime quadrupole radiation formulas, and therefore neglects higher-order post-Newtonian multipole contributions as well as radiation-reaction backreaction on the orbit. Consequently, the waveforms obtained in this work should be regarded as qualitative estimates that capture the dominant strong-field features of the zoom--whirl dynamics, rather than precise templates suitable for matched-filter data analysis pipelines. This level of approximation is standard and well-suited for exploratory studies of the kind presented here \cite{Babak:2006uv, Gair:2005ih}.

The NK procedure begins by integrating the geodesic equations to obtain the time-dependent trajectory $Z^i(t)$ of the test particle in the background spacetime. The resulting position and velocity data are then used to construct the waveform via the standard quadrupole approximation \cite{Maggiore:2007ulw, Misner:1973prb}. The metric perturbation, within this framework, describing the emitted GW is given by
\begin{equation}\label{hij_Iij}
    h_{ij} = \frac{2}{D_L} \, \ddot{I}_{ij}^{\rm STF},
\end{equation}
where $I_{ij}^{\rm STF}$ denotes the symmetric trace-free (STF) mass quadrupole moment, $D_L$ is the luminosity distance to the source, and we adopt natural units $G = c = 1$ (in natural units). The quadrupole moment for an orbiting particle of mass $m$ is defined as
\begin{equation}\label{Iij}
    I_{ij} = m \int d^3x x^i x^j \delta^3\big( x^i - Z^i(t) \big)
\end{equation}
where $Z^i(t)$ shows the position of the moving particle at any time $t$. To compute and interpret the waveform, the choice of the coordinate is very important. We solve the geodesic equations in the spherical coordinate $(r, \theta, \phi)$. On the other hand, it is more convenient to calculate the waveform in the detector-adopted coordinate, which is a Cartesian coordinate $(X, Y, Z)$. This transformation from spherical coordinates to Cartesian coordinates simplifies the analytical calculation of the signal, as measured by the GW detector, and is given by \cite{Babak:2006uv}
\begin{equation}
    x = r \sin\theta \cos\phi, \quad y= r \sin\theta \sin\phi, \quad z = r \cos\theta .
\end{equation}
The coordinate transformation allows the trajectory of the compact object to be expressed in a Cartesian frame centered on the black hole. Substituting the quadrupole moment $I_{ij}=m\,x_i x_j$ into Eq.~\eqref{hij_Iij} and taking two time derivatives, the metric perturbation can be written as \cite{Maggiore:2007ulw, Babak:2006uv}
\begin{equation}
    h_{ij} = \frac{2m}{D_L} 
    \left( a_i x_j + a_j x_i + 2 v_i v_j \right)^{\rm STF},
\end{equation}
where $v_i=\dot{x}_i$ and $a_i=\ddot{x}_i$ denote the Cartesian velocity and acceleration components of the particle, respectively, and STF indicates the symmetric trace-free projection. We continue to work in natural units $G=c=1$. 
The Cartesian coordinate system is centered on the massive black hole. To relate the source frame to the observer’s frame, one performs a rotation characterized by the inclination angle $\iota$, defined as the angle between the line of sight and the orbital angular momentum vector, and the longitude of pericenter $\zeta$ \cite{Babak:2006uv, Maggiore:2007ulw}. This rotation enables the projection of $h_{ij}$ onto the transverse-traceless (TT) gauge and the subsequent extraction of the observable GW polarizations $h_{+}$ and $h_{\times}$. Therefore, the basis of the detector frame in the $(x,y,z)$ coordinates is determined as follows
\begin{eqnarray}
\hat{e}_X &=& (\cos\zeta, -\sin\zeta, 0),\\
\hat{e}_Y &=& (\sin\iota \sin\zeta, \cos\iota \cos\zeta, -\sin\iota),\\
\hat{e}_Z &=& (\sin\iota \sin\zeta, -\sin\iota \cos\zeta, \cos\iota),
\end{eqnarray} 
in which $\hat{e}_X$, $\hat{e}_Y$, and $\hat{e}_Z$ are orthogonal coordinate bases. By projecting the metric perturbation $h_{ij}$ onto the detector frame, one can obtain the two GW polarizations $h_+$ and $h_\times$ as follows
\begin{eqnarray}\label{h_plus_h_cross}
h_+&=\frac{1}{2}\big(e_X^i e_X^j-e_Y^ie_Y^j\big)h_{ij},\\
h_{\times}&=\frac{1}{2}\big(e_X^i e_Y^j-e_Y^ie_X^j\big)h_{ij}.
\end{eqnarray}
Defining the components $h_{\zeta\zeta}$, $h_{\iota\iota}$, and $h_{\iota\zeta}$ in terms of the $h_{ij}$ components as \cite{Babak:2006uv}
\begin{equation}
    h_{\zeta\zeta} = h_{xx}\cos^2\zeta-h_{xy}\sin{2 \zeta}+h_{yy}\sin^2\zeta,
\end{equation}
\begin{eqnarray}
    h_{\iota\iota}&=& \cos^2\iota\big[h_{xx}\sin^2\zeta + h_{xy}\sin 2 \zeta + h_{yy}\cos^2 \zeta\big] \nonumber \\
&& +h_{zz} \sin^2\iota - \sin{2 \iota}\big[h_{xz} \sin\zeta + h_{yz}\cos\zeta\big],
\end{eqnarray}
\begin{eqnarray}
    h_{\iota\zeta}&=& \frac{1}{2}\cos\iota\big[h_{xx} \sin {2 \zeta}+ 2 h_{xy}\cos{2 \zeta}- h_{yy}\sin{2 \zeta}\big]\nonumber\\
    & &+\sin\iota\big[h_{yz} \sin\zeta-h_{xx} \cos\zeta\big],
\end{eqnarray}
the polarizations can be rewritten in terms of these new components as 
\begin{eqnarray}\label{hplus_hcross}
h_+&=&\frac{1}{2}\big(h_{\zeta\zeta}-h_{\iota\iota}\big),\\
h_{\times}&=&h_{\iota\zeta}.
\end{eqnarray}
We analyze the GWs emitted by the smaller compact object as it follows periodic geodesic orbits around the black hole. For definiteness, we consider an EMRI system with component masses $m = 10\,M_\odot$ and $M = 10^{6}\,M_\odot$. The inclination and longitude of pericenter are fixed at $\iota = \zeta = \pi/4$, and the luminosity distance is chosen as $D_L = 200\,\mathrm{Mpc}$. The resulting GW polarizations, $h_{+}$ and $h_{\times}$, are shown in Fig.~\ref{gw1} for representative periodic configurations characterized by $(z,w,v)=(1,2,0)$, $(2,1,1)$, and $(3,2,2)$. Fig.~\ref{gw1} clearly demonstrates the characteristic zoom--whirl structure imprinted in the waveform, a well-known strong-field feature of EMRIs \cite{Levin:2008mq, Barack:2003fp, Babak:2006uv, Gair:2004iv}. The waveform morphology directly reflects the orbital motion over one radial cycle. During the zoom phase, the compact object moves along a highly eccentric, large-radius segment of the orbit, producing relatively smooth, low-amplitude oscillations in the gravitational signal. This phase is intermittently interrupted by short-duration, high-amplitude bursts associated with the whirl phase, during which the particle executes multiple rapid, nearly circular revolutions close to the black hole. The enhanced orbital frequency and strong-field curvature effects during the whirl motion result in prominent high-frequency oscillations in the waveform. Consequently, each orbital cycle includes extended low-amplitude segments corresponding to the zoom motion and sharp high-amplitude bursts corresponding to the whirl motion. Importantly, the number of burst structures in the waveform matches the number of whirls $w$ in the corresponding periodic orbit, consistent with the Levin--Perez-Giz classification \cite{Levin:2008mq}. 


To study the influence of quantum effects on the particle orbits and the waveform signal, we have considered periodic orbits and waveforms for a specific orbit $(z,w,v) = (3,2,2)$ and different values of $\xi = 0, 0.2$, and $0.4$, so that we also included the Schwarzschild case to have a better comparison. The numerical results are plotted in Fig.~\ref{orgw}, where the left panel illustrates the periodic orbits and the right panel displays the numerical results for $h_+$ and $h_\times$. It is observed that the scaling parameter significantly affects the orbits of the moving smaller object around the supermassive black hole and the generated waveform. The waveform amplitude increases with increasing scaling parameter, and a discernible phase shift is observed as this parameter varies\footnote{Although the corrections induced by the parameter $\xi$ are perturbatively small in the strong-field region, GW observables, particularly the phase, are sensitive to cumulative effects over many orbital cycles. As a result, even small fractional shifts in orbital frequencies can lead to potentially measurable imprints in EMRI.}   

To gain a better understanding and clearer insight, the $(1,1,0)$ orbit and its corresponding waveform are plotted in Fig. \ref{orgw1} in different colors. Therefore, one can clearly understand which part of the waveform is related to which part of the period orbit.   

\begin{figure}[ht!] 
\centering
\begin{tabular}{c c c}
	\includegraphics[width=0.45\textwidth]{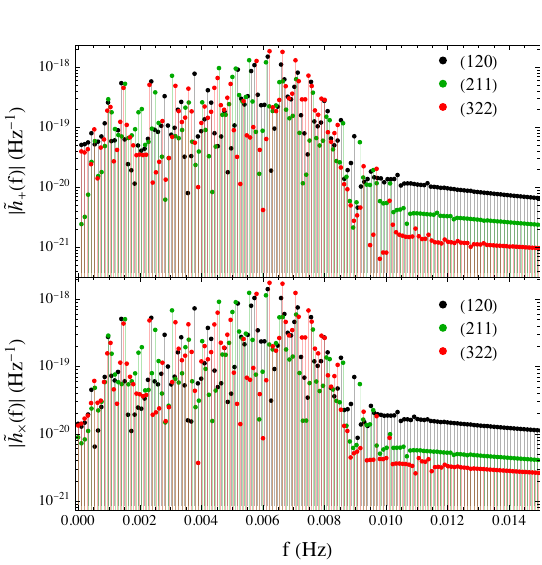}\\
    \end{tabular}
    \caption{Frequency-domain amplitude spectra $|\tilde{h}+(f)|$ and $|\tilde{h}\times(f)|$ for the waveforms shown in Fig. \ref{gw1}, comparing the spectral features of different periodic orbit classifications.}
\label{FRH1}
\end{figure}

To assess the detectability of the emitted signals, we compare our results with the projected sensitivities of future space-based GW detectors such as LISA, Taiji, and TianQin \cite{LISA:2017pwj,Hu:2017mde,TianQin:2015yph}. For this purpose, we compute the frequency-domain waveform components $|\tilde{h}_{+}(f)|$ and $|\tilde{h}_{\times}(f)|$, as well as the corresponding characteristic strain $h_c(f)$, which is the standard quantity used for detector comparison \cite{Maggiore:2007ulw}. The characteristic strain is defined as
\begin{equation}\label{eq:characteristic_strain}
    h_c(f) = 2 f \sqrt{|\tilde{h}_{+}(f)|^2 + |\tilde{h}_{\times}(f)|^2}.
\end{equation}
\begin{figure}[ht!] 
\centering
\begin{tabular}{c c c}
	\includegraphics[width=0.45\textwidth]{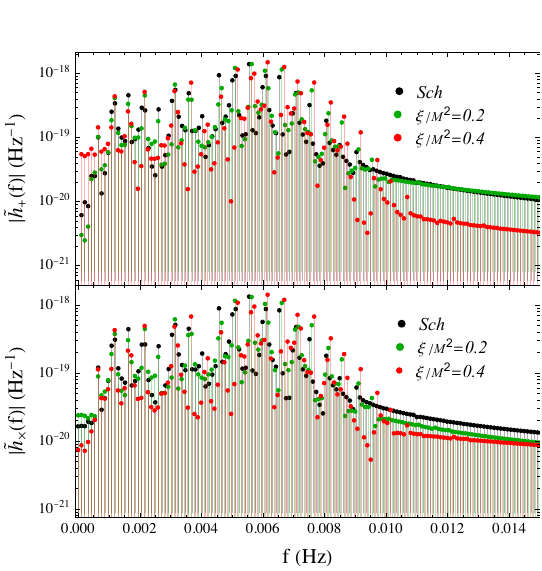}
    \end{tabular}
    \caption{Frequency-domain amplitude spectra $|\tilde{h}+(f)|$ and $|\tilde{h}_{\times}(f)|$ obtained via discrete Fourier transform of the waveforms shown in Fig. \ref{orgw}, illustrating the effect of the quantum parameter $\xi$ on the spectral content.}
\label{FRH}
\end{figure}
The frequency-domain waveform is obtained by executing a discrete Fourier transform of the time-domain signal. This procedure allows us to resolve the spectral content of the gravitational radiation and to identify the dominant harmonic structure associated with periodic zoom--whirl motion. The resulting spectra are shown in Fig.~\ref{FRH} for the orbit $(3,2,2)$ with different values of $\xi$, and in Fig.~\ref{FRH1} for different periodic configurations at fixed $\xi=0.2$.
From these figures, it is evident that the dominant spectral components lie in the millihertz frequency band. This is consistent with the characteristic orbital frequency scale $f \sim (M/10^{6}M_\odot)^{-1}\,\mathrm{mHz}$ expected for EMRIs around supermassive black holes. Therefore, the predicted signals fall within the optimal sensitivity windows of LISA, Taiji, and TianQin, indicating that the present model could be astrophysically relevant for the future space-based GW observations. 

It is important to note that comparing the characteristic strain with detector sensitivity curves only gives a rough idea of detectability. A proper assessment requires a full signal-to-noise ratio (SNR) calculation, which includes the observation time, averaging over source position and orientation, and the detailed noise model of each detector \cite{LISA:2017pwj,Hu:2017mde,TianQin:2015yph}. Such a detailed analysis is beyond the scope of this work. Therefore, our results should be understood as showing the possible qualitative sensitivity of future EMRI observations to the quantum parameter $\xi$, rather than providing firm detection predictions.


\begin{figure}[ht!] 
    \centering
	\includegraphics[width=0.45\textwidth]{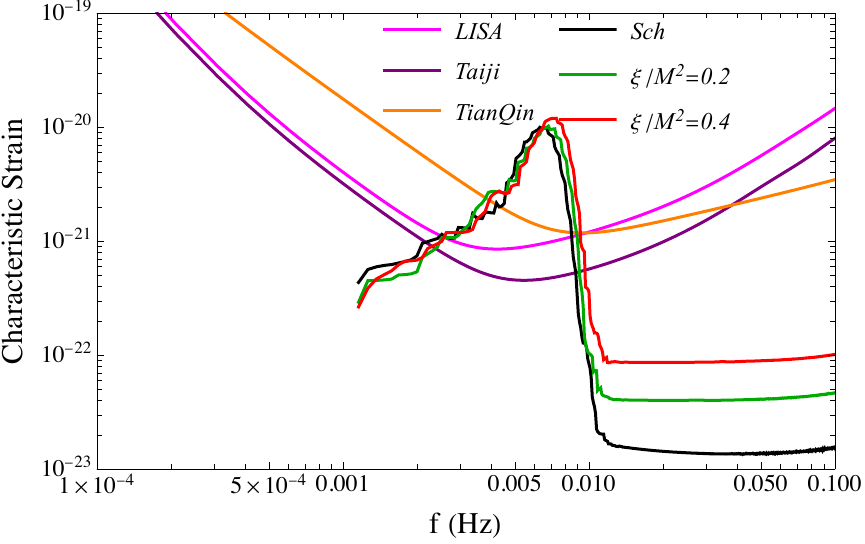}
    \caption{Characteristic strain $h_c(f)$ for the $(3,2,2)$ periodic orbit as a function of frequency for different values of $\xi$: Schwarzschild case $\xi = 0$ (black solid line), $\xi = 0.2M^2$ (green solid line), and $\xi = 0.4M^2$ (red solid line). The sensitivity curves of future space-based GW detectors LISA, Taiji, and TianQin are overlaid for comparison. The signal peaks in the millihertz band and crosses all three detector sensitivity thresholds, with peak amplitude increasing monotonically with $\xi$.}
\label{strain}
\end{figure}

Having obtained the numerical frequency-domain polarizations $\tilde{h}_{+}(f)$ and $\tilde{h}_{\times}(f)$, we compute the corresponding characteristic strain using Eq.~\eqref{eq:characteristic_strain}. The resulting characteristic strain as a function of frequency is shown in Fig.~\ref{strain} for the periodic orbit $(3,2,2)$ and for different values of the scale parameter $\xi$. The solid black curve corresponds to the Schwarzschild case ($\xi=0$), while the green and red solid curves represent $\xi=0.2$ and $\xi=0.4$, respectively. In all cases, the dominant peak of the signal lies in the millihertz frequency band, which overlaps with the projected sensitivity windows of space-based detectors such as LISA, Taiji, and TianQin, indicating that GWs emitted by EMRIs are potentially detectable. Additionally, the strain amplitude is further enhanced as the scale parameter $\xi$ increases, and the peak will shift slightly toward higher frequency as well. These results show the effect of the quantum corrections in ASG on the GW signal from the periodic orbits. With advances in technology, there is a strong prospect of detecting such a GW signal, which would provide a valuable opportunity to study the spacetime geometry of the supermassive black hole and to examine the effects of quantum gravity.    
\begin{figure}[ht!] 
    \centering
	\includegraphics[width=0.45\textwidth]{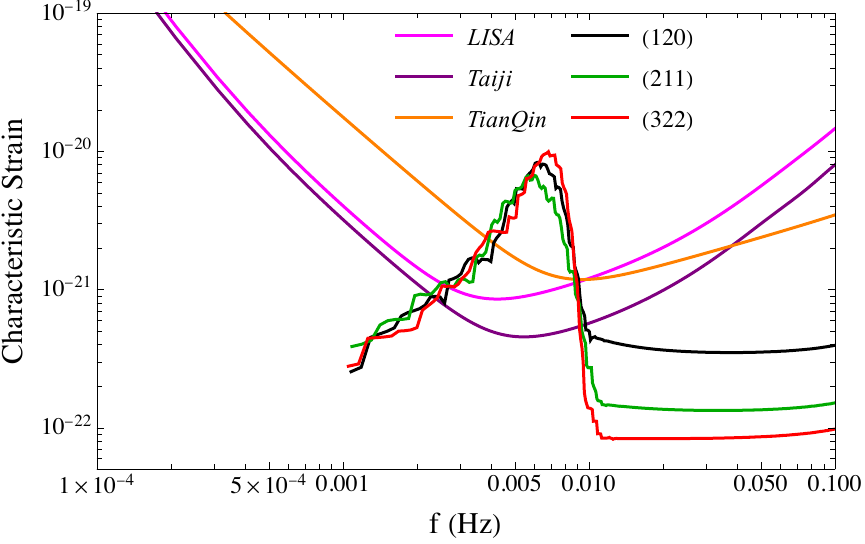}
    \caption{Characteristic strain $h_c(f)$ for different periodic orbit classifications $(z,w,v)$ with fixed $\xi = 0.2M^2$. The sensitivity curves of LISA, Taiji, and TianQin are also shown. The frequency content and strain amplitude vary significantly with the orbit topology, demonstrating how different zoom-whirl structures imprint characteristic features in the detectable GW signal.}
\label{strain1}
\end{figure}

\section{Conclusion}\label{sec:conclusion}
The strong-gravity region near a black hole remains one of the most promising places to test the foundations of gravitational physics. Although GR has passed many experimental tests, its behavior in the near-horizon regime, especially in the presence of possible quantum corrections, remains an open question. The EMRIs are particularly suitable systems for this purpose, since the long-lived orbital motion of a compact object around a supermassive black hole encodes detailed information about the background spacetime in the emitted GWs. Motivated by the anticipated observations from future space-based detectors, it is therefore worthwhile to investigate how quantum-modified RBHASG geometries affect orbital dynamics and gravitational-wave signals.

Indeed, we have investigated the dynamics of a test particle on bound time-like geodesics around the RBHASG metric (\ref{metric}). The spacetime metric of this regular black hole, recently proposed by Bonanno et al. \cite{Bonanno:2023rzk}, arises from the gravitational collapse of dust in ASG. The metric includes a logarithmic term that encodes quantum corrections through the scaling parameter $\xi$, and reduces to the Schwarzschild metric as $\xi$ approaches zero. We focused on periodic orbits of a stellar-mass object (mass $\sim 10 M_\odot$) around a supermassive black hole (mass $\sim 10^6 M_\odot$), their classification using the zoom-whirl taxonomy, and the associated GW emission.
   
Using the introduced static, spherically symmetric metric, the equations of motion for time-like geodesics were first obtained, and the corresponding effective potential was constructed. Solving the required conditions for ISCOs, we found the minimum energy for the bound orbit of the smaller object. Subsequently, we solved the required conditions for MBOs.  
The result indicates that increasing the scaling parameter $\xi$ decreases the radii and angular momenta of these two orbits. Additionally, the energy of the ISCOs decreases with increasing $\xi$. 

The periodic orbits are characterized using the ratio of radial to angular frequencies, in which they are classified in terms of the topological integers $(z,w,v)$. The orbits and the corresponding angular momenta are numerically obtained for different values of the zoom $z$, whirl $w$, and vertex $v$ numbers. It was shown that a more complex orbit is predicted as the zoom number $z$ increases. Then, we computed the GW signal generated by selected orbits using the quadrupole approximation. The two polarization modes, i.e., $h_+$ and $h_\times$, are calculated from the trajectory of the test particle. The time evolution of the polarizations clearly exhibits the zoom and whirl patterns. By applying a discrete Fourier transform to the time-domain signal, we evaluated the frequency-domain waveform and calculated the characteristic strain. Comparing the resulting characteristic strain with the sensitivity curves of LISA, Taiji, and TianQin indicates that the peak reaches the sensitivity limits of all three detectors. However, these results provide a qualitative indication that future space-based detectors could place constraints on $\xi$, with a complete detectability assessment requiring a dedicated signal-to-noise ratio analysis.

Additionally, we examined the influence of the scaling parameter $\xi$ on a representative periodic orbit with $(z,w,v)=(3,2,2)$. The results indicate that varying $\xi$ modifies the orbital dynamics of the smaller object around the supermassive black hole, which in turn leads to noticeable changes in the emitted waveform. In particular, the waveform amplitude increases as the parameter $\xi$ increases. 
Moreover, a cumulative phase shift is observed as $\xi$ varies, with larger values of $\xi$ causing the waveform to advance relative to the Schwarzschild case. This phase difference arises from the modification of the orbital frequencies induced by the quantum correction parameter. The analysis of the corresponding characteristic strain further shows that the peak amplitude grows with increasing $\xi$, accompanied by a slight shift in its frequency location. 

Several extensions of this work are possible. A clear next step would be to go beyond the geodesic and quadrupole approximations by including radiation reaction and self-force effects. This would allow for more realistic inspiral waveforms, which are essential for constructing templates to search for signals in LISA data.
It would also be worthwhile to introduce spin into the RBHASG framework. The spin produces frame-dragging, and its interplay with the quantum parameter $\xi$ could leave distinctive imprints on both orbital dynamics and waveforms. Similarly, moving beyond equatorial motion to study three-dimensional inclined orbits might reveal additional features in the GW signal and expand the range of testable scenarios.
From an observational perspective, applying Bayesian parameter estimation to simulated EMRI events would help clarify how tightly $\xi$ could be constrained with future detectors. Comparing ASG predictions with other quantum-gravity-motivated models—such as those from loop quantum gravity or string theory—could also help identify unique signatures that might distinguish between different theoretical ideas.
As space-based GW observatories become operational, 
refining these theoretical predictions will be key. If quantum effects do leave a mark in the strong-field regime, EMRI observations may provide an opportunity to constrain their effects. 

\bibliographystyle{apsrev4-1}
\bibliography{main.bib}

@article{Wei:2025qlh,
    author = "Wei, Ze-Lin and Zhang, Jing and Xie, Yi and Yin, Pei-Lin",
    title = "{Probing a one-loop quantum-corrected Schwarzschild spacetime with precessing and periodic motion}",
    doi = "10.1140/epjc/s10052-025-14437-x",
    journal = "Eur. Phys. J. C",
    volume = "85",
    number = "6",
    pages = "698",
    year = "2025"
}

@article{Bonanno:2019ilz,
    author = "Bonanno, Alfio and Casadio, Roberto and Platania, Alessia",
    title = "{Gravitational antiscreening in stellar interiors}",
    eprint = "1910.11393",
    archivePrefix = "arXiv",
    primaryClass = "gr-qc",
    doi = "10.1088/1475-7516/2020/01/022",
    journal = "JCAP",
    volume = "01",
    pages = "022",
    year = "2020"
}

@article{Bonanno:2021squ,
    author = "Bonanno, Alfio and Denz, Tobias and Pawlowski, Jan M. and Reichert, Manuel",
    title = "{Reconstructing the graviton}",
    eprint = "2102.02217",
    archivePrefix = "arXiv",
    primaryClass = "hep-th",
    doi = "10.21468/SciPostPhys.12.1.001",
    journal = "SciPost Phys.",
    volume = "12",
    number = "1",
    pages = "001",
    year = "2022"
}

@article{Bonanno:2023rzk,
    author = "Bonanno, Alfio and Malafarina, Daniele and Panassiti, Antonio",
    title = "{Dust Collapse in Asymptotic Safety: A Path to Regular Black Holes}",
    eprint = "2308.10890",
    archivePrefix = "arXiv",
    primaryClass = "gr-qc",
    doi = "10.1103/PhysRevLett.132.031401",
    journal = "Phys. Rev. Lett.",
    volume = "132",
    number = "3",
    pages = "031401",
    year = "2024"
}

@book{Wald:1984rg,
    author = "Wald, Robert M.",
    title = "{General Relativity}",
    doi = "10.7208/chicago/9780226870373.001.0001",
    publisher = "Chicago Univ. Pr.",
    address = "Chicago, USA",
    year = "1984"
}

@book{Shapiro:1983du,
    author = "Shapiro, S. L. and Teukolsky, S. A.",
    title = "{Black holes, white dwarfs, and neutron stars: The physics of compact objects}",
    doi = "10.1002/9783527617661",
    isbn = "978-0-471-87316-7, 978-3-527-61766-1",
    year = "1983"
}

@article{Bardeen:1972fi,
    author = "Bardeen, James M. and Press, William H. and Teukolsky, Saul A",
    title = "{Rotating black holes: Locally nonrotating frames, energy extraction, and scalar synchrotron radiation}",
    reportNumber = "OAP-288",
    doi = "10.1086/151796",
    journal = "Astrophys. J.",
    volume = "178",
    pages = "347",
    year = "1972"
}

@article{Glampedakis:2002cb,
    author = "Glampedakis, Kostas and Hughes, Scott A. and Kennefick, Daniel",
    title = "{Approximating the inspiral of test bodies into Kerr black holes}",
    eprint = "gr-qc/0205033",
    archivePrefix = "arXiv",
    doi = "10.1103/PhysRevD.66.064005",
    journal = "Phys. Rev. D",
    volume = "66",
    pages = "064005",
    year = "2002"
}

@article{Yang:2024cnd,
    author = "Yang, Sen and Zhang, Yu-Peng and Zhu, Tao and Zhao, Li and Liu, Yu-Xiao",
    title = "{Constraining polymerized black holes with quasi-circular extreme mass-ratio inspirals*}",
    eprint = "2412.04302",
    archivePrefix = "arXiv",
    primaryClass = "gr-qc",
    doi = "10.1088/1674-1137/adef1a",
    journal = "Chin. Phys.",
    volume = "49",
    number = "11",
    pages = "115107",
    year = "2025"
}

@book{Misner:1973prb,
    author = "Misner, Charles W. and Thorne, K. S. and Wheeler, J. A.",
    title = "{Gravitation}",
    isbn = "978-0-7167-0344-0, 978-0-691-17779-3",
    publisher = "W. H. Freeman",
    address = "San Francisco",
    year = "1973"
}

@article{Mino:1997bx,
    author = "Mino, Yasushi and Sasaki, Misao and Shibata, Masaru and Tagoshi, Hideyuki and Tanaka, Takahiro",
    title = "{Black hole perturbation: Chapter 1}",
    eprint = "gr-qc/9712057",
    archivePrefix = "arXiv",
    doi = "10.1143/PTPS.128.1",
    journal = "Prog. Theor. Phys. Suppl.",
    volume = "128",
    pages = "1--121",
    year = "1997"
}

@article{Barack:2009ux,
    author = "Barack, Leor",
    title = "{Gravitational self force in extreme mass-ratio inspirals}",
    eprint = "0908.1664",
    archivePrefix = "arXiv",
    primaryClass = "gr-qc",
    doi = "10.1088/0264-9381/26/21/213001",
    journal = "Class. Quant. Grav.",
    volume = "26",
    pages = "213001",
    year = "2009"
}

@article{Gair:2005ih,
    author = "Gair, Jonathan R and Glampedakis, Kostas",
    title = "{Improved approximate inspirals of test-bodies into Kerr black holes}",
    eprint = "gr-qc/0510129",
    archivePrefix = "arXiv",
    doi = "10.1103/PhysRevD.73.064037",
    journal = "Phys. Rev. D",
    volume = "73",
    pages = "064037",
    year = "2006"
}

@book{Maggiore:2007ulw,
    author = "Maggiore, Michele",
    title = "{Gravitational Waves. Vol. 1: Theory and Experiments}",
    doi = "10.1093/acprof:oso/9780198570745.001.0001",
    isbn = "978-0-19-171766-6, 978-0-19-852074-0",
    publisher = "Oxford University Press",
    year = "2007"
}

@article{Gair:2012nm,
    author = "Gair, Jonathan R. and Vallisneri, Michele and Larson, Shane L. and Baker, John G.",
    title = "{Testing General Relativity with Low-Frequency, Space-Based Gravitational-Wave Detectors}",
    eprint = "1212.5575",
    archivePrefix = "arXiv",
    primaryClass = "gr-qc",
    doi = "10.12942/lrr-2013-7",
    journal = "Living Rev. Rel.",
    volume = "16",
    pages = "7",
    year = "2013"
}

@article{Poisson:2011nh,
    author = "Poisson, Eric and Pound, Adam and Vega, Ian",
    title = "{The Motion of point particles in curved spacetime}",
    eprint = "1102.0529",
    archivePrefix = "arXiv",
    primaryClass = "gr-qc",
    doi = "10.12942/lrr-2011-7",
    journal = "Living Rev. Rel.",
    volume = "14",
    pages = "7",
    year = "2011"
}

@article{Zahra:2025tdo,
    author = "Zahra, Tehreem and Shabbir, Oreeda and Majeed, Bushra and Jamil, Mubasher and Rayimbaev, Javlon and Shermatov, Abubakir",
    title = "{Gravitational wave radiation from periodic orbits and quasi-periodic oscillations in Einstein non-linear Maxwell-Yukawa black hole}",
    eprint = "2510.22761",
    archivePrefix = "arXiv",
    primaryClass = "gr-qc",
    month = "10",
    year = "2025"
}

@article{Li:2025eln,
    author = "Li, Guo-He and Qiao, Chen-Kai and Tao, Jun",
    title = "{Periodic orbits and their gravitational waves in EMRIs: supermassive black hole affected by galactic dark matter halos}",
    eprint = "2510.24989",
    archivePrefix = "arXiv",
    primaryClass = "gr-qc",
    month = "10",
    year = "2025"
}

@article{Chen:2025aqh,
    author = "Chen, Jiawei and Yang, Jinsong",
    title = "{Periodic orbits and gravitational waveforms in quantum-corrected black hole spacetimes}",
    eprint = "2505.02660",
    archivePrefix = "arXiv",
    primaryClass = "gr-qc",
    doi = "10.1140/epjc/s10052-025-14457-7",
    journal = "Eur. Phys. J. C",
    volume = "85",
    number = "7",
    pages = "726",
    year = "2025"
}

@article{Deng:2025wzz,
    author = "Deng, Weike and Long, Sheng and Tan, Qin and Jing, Jiliang",
    title = "{Gravitational waveforms from periodic orbits around a charged black hole with scalar hair}",
    eprint = "2510.24468",
    archivePrefix = "arXiv",
    primaryClass = "gr-qc",
    month = "10",
    year = "2025"
}

@article{Choudhury:2025qsh,
    author = "Choudhury, Sayantan and Hossain, Md Khalid and Bauyrzhan, Gulnur and Yerzhanov, Koblandy",
    title = "{Gravitational Wave Signatures of Periodic Motion near Higher-Derivative Einstein-{\AE}ther Black Holes}",
    eprint = "2507.00904",
    archivePrefix = "arXiv",
    primaryClass = "gr-qc",
    month = "7",
    year = "2025"
}

@article{Li:2025sfe,
    author = "Li, Yong-Zhuang and Kuang, Xiao-Mei",
    title = "{The bound orbits and gravitational waveforms of timelike particles around renormalization group improved Kerr black holes}",
    eprint = "2509.07333",
    archivePrefix = "arXiv",
    primaryClass = "gr-qc",
    month = "9",
    year = "2025"
}

@article{Alloqulov:2025bxh,
    author = "Alloqulov, Mirzabek and Shaymatov, Sanjar and Ahmedov, Bobomurat and Zhu, Tao",
    title = "{Regular black hole's impact on the gravitational waveforms from periodic orbits}",
    eprint = "2508.05245",
    archivePrefix = "arXiv",
    primaryClass = "gr-qc",
    month = "8",
    year = "2025"
}

@article{Gong:2025mne,
    author = "Gong, Huajie and Long, Sheng and Wang, Xi-Jing and Xia, Zhongwu and Wu, Jian-Pin and Pan, Qiyuan",
    title = "{Gravitational waveforms from periodic orbits around a novel regular black hole}",
    eprint = "2509.23318",
    archivePrefix = "arXiv",
    primaryClass = "gr-qc",
    month = "9",
    year = "2025"
}

@article{Zare:2025aek,
    author = "Zare, Soroush and Zhu, Tao and Nieto, Luis M. and Lu, Shuo and Hassanabadi, Hassan",
    title = "{Probing regular black holes with sub-Planckian curvature through periodic orbits and their gravitational wave radiation}",
    eprint = "2510.05166",
    archivePrefix = "arXiv",
    primaryClass = "gr-qc",
    month = "10",
    year = "2025"
}

@article{Lu:2025cxx,
    author = "Lu, Shuo and Zhu, Tao",
    title = "{Gravitational radiations from periodic orbits around Einstein-{\AE}ther black holes}",
    eprint = "2505.00294",
    archivePrefix = "arXiv",
    primaryClass = "gr-qc",
    doi = "10.1016/j.dark.2025.102141",
    journal = "Phys. Dark Univ.",
    volume = "50",
    pages = "102141",
    year = "2025"
}

@article{LIGOScientific:2016aoc,
    author = "Abbott, B. P. and others",
    collaboration = "LIGO Scientific, Virgo",
    title = "{Observation of Gravitational Waves from a Binary Black Hole Merger}",
    eprint = "1602.03837",
    archivePrefix = "arXiv",
    primaryClass = "gr-qc",
    reportNumber = "LIGO-P150914",
    doi = "10.1103/PhysRevLett.116.061102",
    journal = "Phys. Rev. Lett.",
    volume = "116",
    number = "6",
    pages = "061102",
    year = "2016"
}

@article{LIGOScientific:2016vbw,
    author = "Abbott, B. P. and others",
    collaboration = "LIGO Scientific, Virgo",
    title = "{GW150914: First results from the search for binary black hole coalescence with Advanced LIGO}",
    eprint = "1602.03839",
    archivePrefix = "arXiv",
    primaryClass = "gr-qc",
    reportNumber = "LIGO-P1500269",
    doi = "10.1103/PhysRevD.93.122003",
    journal = "Phys. Rev. D",
    volume = "93",
    number = "12",
    pages = "122003",
    year = "2016"
}

@article{LIGOScientific:2016vlm,
    author = "Abbott, B. P. and others",
    collaboration = "LIGO Scientific, Virgo",
    title = "{Properties of the Binary Black Hole Merger GW150914}",
    eprint = "1602.03840",
    archivePrefix = "arXiv",
    primaryClass = "gr-qc",
    reportNumber = "LIGO-P1500218",
    doi = "10.1103/PhysRevLett.116.241102",
    journal = "Phys. Rev. Lett.",
    volume = "116",
    number = "24",
    pages = "241102",
    year = "2016"
}

@article{LIGOScientific:2016emj,
    author = "Abbott, B. P. and others",
    collaboration = "LIGO Scientific, Virgo",
    title = "{GW150914: The Advanced LIGO Detectors in the Era of First Discoveries}",
    eprint = "1602.03838",
    archivePrefix = "arXiv",
    primaryClass = "gr-qc",
    reportNumber = "LIGO-P1500237",
    doi = "10.1103/PhysRevLett.116.131103",
    journal = "Phys. Rev. Lett.",
    volume = "116",
    number = "13",
    pages = "131103",
    year = "2016"
}

@article{Levin:2008mq,
    author = "Levin, Janna and Perez-Giz, Gabe",
    title = "{A Periodic Table for Black Hole Orbits}",
    eprint = "0802.0459",
    archivePrefix = "arXiv",
    primaryClass = "gr-qc",
    doi = "10.1103/PhysRevD.77.103005",
    journal = "Phys. Rev. D",
    volume = "77",
    pages = "103005",
    year = "2008"
}

@article{Levin:2009sk,
    author = "Levin, Janna",
    title = "{Energy Level Diagrams for Black Hole Orbits}",
    eprint = "0907.5195",
    archivePrefix = "arXiv",
    primaryClass = "gr-qc",
    doi = "10.1088/0264-9381/26/23/235010",
    journal = "Class. Quant. Grav.",
    volume = "26",
    pages = "235010",
    year = "2009"
}

@article{Misra:2010pu,
    author = "Misra, Vedant and Levin, Janna",
    title = "{Rational Orbits around Charged Black Holes}",
    eprint = "1007.2699",
    archivePrefix = "arXiv",
    primaryClass = "gr-qc",
    doi = "10.1103/PhysRevD.82.083001",
    journal = "Phys. Rev. D",
    volume = "82",
    pages = "083001",
    year = "2010"
}

@article{Babar:2017gsg,
    author = "Babar, Gulmina Zaman and Babar, Adil Zaman and Lim, Yen-Kheng",
    title = "{Periodic orbits around a spherically symmetric naked singularity}",
    eprint = "1710.09581",
    archivePrefix = "arXiv",
    primaryClass = "gr-qc",
    doi = "10.1103/PhysRevD.96.084052",
    journal = "Phys. Rev. D",
    volume = "96",
    number = "8",
    pages = "084052",
    year = "2017"
}

@article{Hu:2017mde,
    author = "Hu, Wen-Rui and Wu, Yue-Liang",
    title = "{The Taiji Program in Space for gravitational wave physics and the nature of gravity}",
    doi = "10.1093/nsr/nwx116",
    journal = "Natl. Sci. Rev.",
    volume = "4",
    number = "5",
    pages = "685--686",
    year = "2017"
}

@article{TianQin:2015yph,
    author = "Luo, Jun and others",
    collaboration = "TianQin",
    title = "{TianQin: a space-borne gravitational wave detector}",
    eprint = "1512.02076",
    archivePrefix = "arXiv",
    primaryClass = "astro-ph.IM",
    doi = "10.1088/0264-9381/33/3/035010",
    journal = "Class. Quant. Grav.",
    volume = "33",
    number = "3",
    pages = "035010",
    year = "2016"
}

@article{Gong:2021gvw,
    author = "Gong, Yungui and Luo, Jun and Wang, Bin",
    title = "{Concepts and status of Chinese space gravitational wave detection projects}",
    eprint = "2109.07442",
    archivePrefix = "arXiv",
    primaryClass = "astro-ph.IM",
    doi = "10.1038/s41550-021-01480-3",
    journal = "Nature Astron.",
    volume = "5",
    number = "9",
    pages = "881--889",
    year = "2021"
}

@article{Danzmann:1997hm,
    author = "Danzmann, K.",
    title = "{LISA: An ESA cornerstone mission for a gravitational wave observatory}",
    doi = "10.1088/0264-9381/14/6/002",
    journal = "Class. Quant. Grav.",
    volume = "14",
    pages = "1399--1404",
    year = "1997"
}

@article{Schutz:1999xj,
    author = "Schutz, Bernard F.",
    title = "{Gravitational wave astronomy}",
    eprint = "gr-qc/9911034",
    archivePrefix = "arXiv",
    reportNumber = "AEI-1999-34",
    doi = "10.1088/0264-9381/16/12A/307",
    journal = "Class. Quant. Grav.",
    volume = "16",
    pages = "A131--A156",
    year = "1999"
}

@article{Gair:2004iv,
    author = "Gair, Jonathan R. and Barack, Leor and Creighton, Teviet and Cutler, Curt and Larson, Shane L. and Phinney, E. Sterl and Vallisneri, Michele",
    title = "{Event rate estimates for LISA extreme mass ratio capture sources}",
    eprint = "gr-qc/0405137",
    archivePrefix = "arXiv",
    doi = "10.1088/0264-9381/21/20/003",
    journal = "Class. Quant. Grav.",
    volume = "21",
    pages = "S1595--S1606",
    year = "2004"
}

@article{LISA:2017pwj,
    author = "Amaro-Seoane, Pau and others",
    collaboration = "LISA",
    title = "{Laser Interferometer Space Antenna}",
    eprint = "1702.00786",
    archivePrefix = "arXiv",
    primaryClass = "astro-ph.IM",
    month = "2",
    year = "2017"
}

@article{Maselli:2021men,
    author = "Maselli, Andrea and Franchini, Nicola and Gualtieri, Leonardo and Sotiriou, Thomas P. and Barsanti, Susanna and Pani, Paolo",
    title = "{Detecting fundamental fields with LISA observations of gravitational waves from extreme mass-ratio inspirals}",
    eprint = "2106.11325",
    archivePrefix = "arXiv",
    primaryClass = "gr-qc",
    doi = "10.1038/s41550-021-01589-5",
    journal = "Nature Astron.",
    volume = "6",
    number = "4",
    pages = "464--470",
    year = "2022"
}

@article{Levin:2008ci,
    author = "Levin, Janna and Grossman, Becky",
    title = "{Dynamics of Black Hole Pairs. I. Periodic Tables}",
    eprint = "0809.3838",
    archivePrefix = "arXiv",
    primaryClass = "gr-qc",
    doi = "10.1103/PhysRevD.79.043016",
    journal = "Phys. Rev. D",
    volume = "79",
    pages = "043016",
    year = "2009"
}

@article{Bambhaniya:2020zno,
    author = "Bambhaniya, Parth and Solanki, Divyesh N. and Dey, Dipanjan and Joshi, Ashok B. and Joshi, Pankaj S. and Patel, Vishva",
    title = "{Precession of timelike bound orbits in Kerr spacetime}",
    eprint = "2007.12086",
    archivePrefix = "arXiv",
    primaryClass = "gr-qc",
    doi = "10.1140/epjc/s10052-021-08997-x",
    journal = "Eur. Phys. J. C",
    volume = "81",
    number = "3",
    pages = "205",
    year = "2021"
}

@article{Rana:2019bsn,
    author = "Rana, Prerna and Mangalam, A.",
    title = "{Astrophysically relevant bound trajectories around a Kerr black hole}",
    eprint = "1901.02730",
    archivePrefix = "arXiv",
    primaryClass = "gr-qc",
    doi = "10.1088/1361-6382/ab004c",
    journal = "Class. Quant. Grav.",
    volume = "36",
    pages = "045009",
    year = "2019"
}

@article{Liu:2018vea,
    author = "Liu, Changqing and Ding, Chikun and Jing, Jiliang",
    title = "{Periodic orbits around Kerr Sen black holes}",
    eprint = "1804.05883",
    archivePrefix = "arXiv",
    primaryClass = "gr-qc",
    doi = "10.1088/0253-6102/71/12/1461",
    journal = "Commun. Theor. Phys.",
    volume = "71",
    number = "12",
    pages = "1461",
    year = "2019"
}

@article{Lin:2023rmo,
    author = "Lin, Hou-Yu and Deng, Xue-Mei",
    title = "{Precessing and periodic orbits around hairy black holes in Horndeski{\textquoteright}s Theory}",
    doi = "10.1140/epjc/s10052-023-11487-x",
    journal = "Eur. Phys. J. C",
    volume = "83",
    number = "4",
    pages = "311",
    year = "2023"
}

@article{Yao:2023ziq,
    author = "Yao, Jin-Tao and Li, Xin",
    title = "{Closed orbits in axial symmetric Finslerian extension of a Schwarzschild black hole}",
    doi = "10.1103/PhysRevD.108.084067",
    journal = "Phys. Rev. D",
    volume = "108",
    number = "8",
    pages = "084067",
    year = "2023"
}

@article{Lin:2022llz,
    author = "Lin, Hou-Yu and Deng, Xue-Mei",
    title = "{Bound Orbits and Epicyclic Motions around Renormalization Group Improved Schwarzschild Black Holes}",
    doi = "10.3390/universe8050278",
    journal = "Universe",
    volume = "8",
    number = "5",
    pages = "278",
    year = "2022"
}

@article{Chan:2025ocy,
    author = "Chan, Zoe C. S. and Lim, Yen-Kheng",
    title = {{Periodic orbits of neutral test particles in Reissner{\textendash}Nordstr{\"o}m naked singularities}},
    eprint = "2502.03082",
    archivePrefix = "arXiv",
    primaryClass = "gr-qc",
    doi = "10.1007/s10714-025-03368-3",
    journal = "Gen. Rel. Grav.",
    volume = "57",
    number = "2",
    pages = "35",
    year = "2025"
}

@article{Wang:2022tfo,
    author = "Wang, Ruifang and Gao, Fabao and Chen, Huixiang",
    title = "{Periodic orbits around a static spherically symmetric black hole surrounded by quintessence}",
    doi = "10.1016/j.aop.2022.169167",
    journal = "Annals Phys.",
    volume = "447",
    number = "1",
    pages = "169167",
    year = "2022"
}

@article{Lin:2023eyd,
    author = "Lin, Hou-Yu and Deng, Xue-Mei",
    title = "{Dynamics of test particles around hairy black holes in Horndeski{\textquoteright}s theory}",
    doi = "10.1016/j.aop.2023.169360",
    journal = "Annals Phys.",
    volume = "455",
    pages = "169360",
    year = "2023"
}

@article{Habibina:2022ztd,
    author = "Habibina, A. S. and Ramadhan, H. S.",
    title = "{Bound orbits around charged black strings}",
    eprint = "2205.14635",
    archivePrefix = "arXiv",
    primaryClass = "gr-qc",
    doi = "10.1016/j.aop.2022.169169",
    journal = "Annals Phys.",
    volume = "448",
    pages = "169169",
    year = "2023"
}

@article{Zhang:2022psr,
    author = "Zhang, Jing and Xie, Yi",
    title = "{Probing a self-complete and Generalized-Uncertainty-Principle black hole with precessing and periodic motion}",
    doi = "10.1007/s10509-022-04046-5",
    journal = "Astrophys. Space Sci.",
    volume = "367",
    number = "2",
    pages = "17",
    year = "2022"
}

@article{Lin:2022wda,
    author = "Lin, Hou-Yu and Deng, Xue-Mei",
    title = "{Precessing and periodic orbits around Lee{\textendash}Wick black holes}",
    doi = "10.1140/epjp/s13360-022-02391-6",
    journal = "Eur. Phys. J. Plus",
    volume = "137",
    number = "2",
    pages = "176",
    year = "2022"
}

@article{Gao:2021arw,
    author = "Gao, Bo and Deng, Xue-Mei",
    title = "{Bound orbits around modified Hayward black holes}",
    doi = "10.1142/S0217732321502370",
    journal = "Mod. Phys. Lett. A",
    volume = "36",
    number = "33",
    pages = "2150237",
    year = "2021"
}

@article{Lin:2021noq,
    author = "Lin, Hou-Yu and Deng, Xue-Mei",
    title = "{Rational orbits around 4 $\mathcal D$ Einstein{\textendash}Lovelock black holes}",
    doi = "10.1016/j.dark.2020.100745",
    journal = "Phys. Dark Univ.",
    volume = "31",
    pages = "100745",
    year = "2021"
}

@article{Deng:2020yfm,
    author = "Deng, Xue-Mei",
    title = "{Geodesics and periodic orbits around quantum-corrected black holes}",
    doi = "10.1016/j.dark.2020.100629",
    journal = "Phys. Dark Univ.",
    volume = "30",
    pages = "100629",
    year = "2020"
}

@article{Zhou:2020zys,
    author = "Zhou, Tian-Yi and Xie, Yi",
    title = "{Precessing and periodic motions around a black-bounce/traversable wormhole}",
    doi = "10.1140/epjc/s10052-020-08661-w",
    journal = "Eur. Phys. J. C",
    volume = "80",
    number = "11",
    pages = "1070",
    year = "2020"
}

@article{Wang:2025wob,
    author = "Wang, Chao-Hui and Zhang, Yu-Peng and Zhu, Tao and Wei, Shao-Wen",
    title = "{A new type of multi-branch periodic orbits in dyonic black holes}",
    eprint = "2508.20558",
    archivePrefix = "arXiv",
    primaryClass = "gr-qc",
    month = "8",
    year = "2025"
}

@article{Gao:2020wjz,
    author = "Gao, Bo and Deng, Xue-Mei",
    title = "{Bound orbits around Bardeen black holes}",
    doi = "10.1016/j.aop.2020.168194",
    journal = "Annals Phys.",
    volume = "418",
    pages = "168194",
    year = "2020"
}

@article{Deng:2020hxw,
    author = "Deng, Xue-Mei",
    title = "{Periodic orbits around brane-world black holes}",
    doi = "10.1140/epjc/s10052-020-8067-7",
    journal = "Eur. Phys. J. C",
    volume = "80",
    number = "6",
    pages = "489",
    year = "2020"
}

@article{Wei:2019zdf,
    author = "Wei, Shao-Wen and Yang, Jie and Liu, Yu-Xiao",
    title = "{Geodesics and periodic orbits in Kehagias-Sfetsos black holes in deformed Hor̆ava-Lifshitz gravity}",
    eprint = "1904.03129",
    archivePrefix = "arXiv",
    primaryClass = "gr-qc",
    doi = "10.1103/PhysRevD.99.104016",
    journal = "Phys. Rev. D",
    volume = "99",
    number = "10",
    pages = "104016",
    year = "2019"
}

@article{Pugliese:2013xfa,
    author = "Pugliese, Daniela and Quevedo, Hernando and Ruffini, Remo",
    title = {{General classification of charged test particle circular orbits in Reissner--Nordstr{\"o}m spacetime}},
    eprint = "1304.2940",
    archivePrefix = "arXiv",
    primaryClass = "gr-qc",
    doi = "10.1140/epjc/s10052-017-4769-x",
    journal = "Eur. Phys. J. C",
    volume = "77",
    number = "4",
    pages = "206",
    year = "2017"
}

@article{Healy:2009zm,
    author = "Healy, James and Levin, Janna and Shoemaker, Deirdre",
    title = "{Zoom-Whirl Orbits in Black Hole Binaries}",
    eprint = "0907.0671",
    archivePrefix = "arXiv",
    primaryClass = "gr-qc",
    doi = "10.1103/PhysRevLett.103.131101",
    journal = "Phys. Rev. Lett.",
    volume = "103",
    pages = "131101",
    year = "2009"
}

@article{Zhang:2022zox,
    author = "Zhang, Jing and Xie, Yi",
    title = {{Probing a black-bounce-Reissner{\textendash}Nordstr{\"o}m spacetime with precessing and periodic motion}},
    doi = "10.1140/epjc/s10052-022-10846-4",
    journal = "Eur. Phys. J. C",
    volume = "82",
    number = "10",
    pages = "854",
    year = "2022"
}

@article{Haroon:2025rzx,
    author = "Haroon, Sumarna and Zhu, Tao",
    title = "{Periodic orbits and their gravitational wave radiations in a black hole with a dark matter halo}",
    eprint = "2502.09171",
    archivePrefix = "arXiv",
    primaryClass = "gr-qc",
    doi = "10.1103/ckdt-wtsl",
    journal = "Phys. Rev. D",
    volume = "112",
    number = "4",
    pages = "044046",
    year = "2025"
}

@article{Tu:2023xab,
    author = "Tu, Ze-Yi and Zhu, Tao and Wang, Anzhong",
    title = "{Periodic orbits and their gravitational wave radiations in a polymer black hole in loop quantum gravity}",
    eprint = "2304.14160",
    archivePrefix = "arXiv",
    primaryClass = "gr-qc",
    doi = "10.1103/PhysRevD.108.024035",
    journal = "Phys. Rev. D",
    volume = "108",
    number = "2",
    pages = "024035",
    year = "2023"
}

@article{Azreg-Ainou:2020bfl,
    author = {Azreg-A{\"\i}nou, Mustapha and Chen, Zihang and Deng, Bojun and Jamil, Mubasher and Zhu, Tao and Wu, Qiang and Lim, Yen-Kheng},
    title = "{Orbital mechanics and quasiperiodic oscillation resonances of black holes in Einstein-{\AE}ther theory}",
    eprint = "2004.02602",
    archivePrefix = "arXiv",
    primaryClass = "gr-qc",
    doi = "10.1103/PhysRevD.102.044028",
    journal = "Phys. Rev. D",
    volume = "102",
    number = "4",
    pages = "044028",
    year = "2020"
}

@article{Yang:2024lmj,
    author = "Yang, Sen and Zhang, Yu-Peng and Zhu, Tao and Zhao, Li and Liu, Yu-Xiao",
    title = "{Gravitational waveforms from periodic orbits around a quantum-corrected black hole}",
    eprint = "2407.00283",
    archivePrefix = "arXiv",
    primaryClass = "gr-qc",
    doi = "10.1088/1475-7516/2025/01/091",
    journal = "JCAP",
    volume = "01",
    pages = "091",
    year = "2025"
}

@article{Shabbir:2025kqh,
    author = {Shabbir, Oreeda and Jamil, Mubasher and Azreg-A{\"\i}nou, Mustapha},
    title = "{Periodic orbits and their gravitational wave radiations around the Schwarzschild-MOG black hole}",
    eprint = "2501.04367",
    archivePrefix = "arXiv",
    primaryClass = "gr-qc",
    doi = "10.1016/j.dark.2025.101816",
    journal = "Phys. Dark Univ.",
    volume = "47",
    pages = "101816",
    year = "2025"
}

@article{Junior:2024tmi,
    author = "Junior, Ednaldo L. B. and Junior, Jos{\'e} Tarciso S. S. and Lobo, Francisco S. N. and Rodrigues, Manuel E. and Rubiera-Garcia, Diego and da Silva, Lu{\'\i}s F. Dias and Vieira, Henrique A.",
    title = "{Periodical orbits and waveforms with spontaneous Lorentz symmetry-breaking in Kalb{\textendash}Ramond gravity}",
    eprint = "2412.00769",
    archivePrefix = "arXiv",
    primaryClass = "gr-qc",
    doi = "10.1140/epjc/s10052-025-14299-3",
    journal = "Eur. Phys. J. C",
    volume = "85",
    number = "5",
    pages = "557",
    year = "2025"
}

@article{Zhao:2024exh,
    author = "Zhao, Lai and Tang, Meirong and Xu, Zhaoyi",
    title = "{Periodic orbits and gravitational wave radiation in short hair black hole spacetimes for an extreme mass ratio system}",
    eprint = "2411.01979",
    archivePrefix = "arXiv",
    primaryClass = "gr-qc",
    doi = "10.1140/epjc/s10052-025-13767-0",
    journal = "Eur. Phys. J. C",
    volume = "85",
    number = "1",
    pages = "36",
    year = "2025"
}

@article{Jiang:2024cpe,
    author = "Jiang, Hanyu and Alloqulov, Mirzabek and Wu, Qiang and Shaymatov, Sanjar and Zhu, Tao",
    title = "{Periodic orbits and plasma effects on gravitational weak lensing by self-dual black hole in loop quantum gravity}",
    doi = "10.1016/j.dark.2024.101627",
    journal = "Phys. Dark Univ.",
    volume = "46",
    pages = "101627",
    year = "2024"
}

@article{Meng:2024cnq,
    author = "Meng, Liping and Xu, Zhaoyi and Tang, Meirong",
    title = "{Bound orbits and gravitational wave radiation around the hairy black hole}",
    eprint = "2411.01858",
    archivePrefix = "arXiv",
    primaryClass = "gr-qc",
    doi = "10.1140/epjc/s10052-025-14032-0",
    journal = "Eur. Phys. J. C",
    volume = "85",
    number = "3",
    pages = "306",
    year = "2025"
}

@article{Li:2024tld,
    author = "Li, Yong-Zhuang and Kuang, Xiao-Mei and Sang, Yu",
    title = "{Precessing and periodic timelike orbits and their potential applications in Einsteinian cubic gravity}",
    eprint = "2401.16071",
    archivePrefix = "arXiv",
    primaryClass = "gr-qc",
    doi = "10.1140/epjc/s10052-024-12895-3",
    journal = "Eur. Phys. J. C",
    volume = "84",
    number = "5",
    pages = "529",
    year = "2024"
}

@article{QiQi:2024dwc,
    author = "Qi, Qi and Kuang, Xiao-Mei and Li, Yong-Zhuang and Sang, Yu",
    title = "{Timelike bound orbits and pericenter precession around black hole with conformally coupled scalar hair}",
    eprint = "2407.01958",
    archivePrefix = "arXiv",
    primaryClass = "gr-qc",
    doi = "10.1140/epjc/s10052-024-12989-y",
    journal = "Eur. Phys. J. C",
    volume = "84",
    number = "6",
    pages = "645",
    year = "2024"
}

@article{Alloqulov:2025ucf,
    author = "Alloqulov, Mirzabek and Xamidov, Tursunali and Shaymatov, Sanjar and Ahmedov, Bobomurat",
    title = "{Gravitational waveforms from periodic orbits around a Schwarzschild black hole embedded in a Dehnen-type dark matter halo}",
    eprint = "2504.05236",
    archivePrefix = "arXiv",
    primaryClass = "gr-qc",
    doi = "10.1140/epjc/s10052-025-14529-8",
    journal = "Eur. Phys. J. C",
    volume = "85",
    number = "7",
    pages = "798",
    year = "2025"
}

@article{Wang:2025hla,
    author = "Wang, Chao-Hui and Meng, Xiang-Cheng and Zhang, Yu-Peng and Zhu, Tao and Wei, Shao-Wen",
    title = "{Equatorial periodic orbits and gravitational waveforms in a black hole free of Cauchy horizon}",
    eprint = "2502.08994",
    archivePrefix = "arXiv",
    primaryClass = "gr-qc",
    doi = "10.1088/1475-7516/2025/07/021",
    month = "2",
    year = "2025"
}

@article{Babak:2006uv,
    author = "Babak, Stanislav and Fang, Hua and Gair, Jonathan R. and Glampedakis, Kostas and Hughes, Scott A.",
    title = "{'Kludge' gravitational waveforms for a test-body orbiting a Kerr black hole}",
    eprint = "gr-qc/0607007",
    archivePrefix = "arXiv",
    doi = "10.1103/PhysRevD.75.024005",
    journal = "Phys. Rev. D",
    volume = "75",
    pages = "024005",
    year = "2007",
    note = "[Erratum: Phys.Rev.D 77, 04990 (2008)]"
}

@book{Chandrasekhar:1985kt,
    author = "Chandrasekhar, Subrahmanyan",
    title = "{The mathematical theory of black holes}",
    isbn = "978-0-19-850370-5",
    year = "1985"
}

@article{Barack:2003fp,
    author = "Barack, Leor and Cutler, Curt",
    title = "{LISA capture sources: Approximate waveforms, signal-to-noise ratios, and parameter estimation accuracy}",
    eprint = "gr-qc/0310125",
    archivePrefix = "arXiv",
    doi = "10.1103/PhysRevD.69.082005",
    journal = "Phys. Rev. D",
    volume = "69",
    pages = "082005",
    year = "2004"
}

@article{Ahmed:2025azu,
    author = "Ahmed, Fazlay and Wu, Qiang and Ghosh, Sushant G. and Zhu, Tao",
    title = "{Gravitational Wave Signatures from Periodic Orbits around a non-commutative inspired black hole surrounded by quintessence}",
    eprint = "2511.08456",
    archivePrefix = "arXiv",
    primaryClass = "gr-qc",
    month = "11",
    year = "2025"
}

@article{EventHorizonTelescope:2019dse,
    author = "Akiyama, Kazunori and others",
    collaboration = "Event Horizon Telescope",
    title = "{First M87 Event Horizon Telescope Results. I. The Shadow of the Supermassive Black Hole}",
    eprint = "1906.11238",
    archivePrefix = "arXiv",
    primaryClass = "astro-ph.GA",
    doi = "10.3847/2041-8213/ab0ec7",
    journal = "Astrophys. J. Lett.",
    volume = "875",
    pages = "L1",
    year = "2019"
}

@article{EventHorizonTelescope:2022wkp,
    author = "Akiyama, Kazunori and others",
    collaboration = "Event Horizon Telescope",
    title = "{First Sagittarius A* Event Horizon Telescope Results. I. The Shadow of the Supermassive Black Hole in the Center of the Milky Way}",
    eprint = "2311.08680",
    archivePrefix = "arXiv",
    primaryClass = "astro-ph.HE",
    doi = "10.3847/2041-8213/ac6674",
    journal = "Astrophys. J. Lett.",
    volume = "930",
    number = "2",
    pages = "L12",
    year = "2022"
}

@article{Bardeen:1968proceeding,
    author = "Bardeen, J.",
    title = "Non-singular general relativistic gravitational collapse",
    journal = "in Proceedings of the 5th International Conference on
Gravitation and the Theory of Relativity", 
    page = "p.87",
    year = "1968"
}

@article{Hayward:2005gi,
    author = "Hayward, Sean A.",
    title = "{Formation and evaporation of regular black holes}",
    eprint = "gr-qc/0506126",
    archivePrefix = "arXiv",
    doi = "10.1103/PhysRevLett.96.031103",
    journal = "Phys. Rev. Lett.",
    volume = "96",
    pages = "031103",
    year = "2006"
}

@article{Bronnikov:2005gm,
    author = "Bronnikov, K. A. and Fabris, J. C.",
    title = "{Regular phantom black holes}",
    eprint = "gr-qc/0511109",
    archivePrefix = "arXiv",
    doi = "10.1103/PhysRevLett.96.251101",
    journal = "Phys. Rev. Lett.",
    volume = "96",
    pages = "251101",
    year = "2006"
}

@article{Burinskii:2001bq,
    author = "Burinskii, Alexander and Elizalde, Emilio and Hildebrandt, Sergi R. and Magli, Giulio",
    title = "{Regular sources of the Kerr-Schild class for rotating and nonrotating black hole solutions}",
    eprint = "gr-qc/0109085",
    archivePrefix = "arXiv",
    doi = "10.1103/PhysRevD.65.064039",
    journal = "Phys. Rev. D",
    volume = "65",
    pages = "064039",
    year = "2002"
}

@article{Fan:2016hvf,
    author = "Fan, Zhong-Ying and Wang, Xiaobao",
    title = "{Construction of Regular Black Holes in General Relativity}",
    eprint = "1610.02636",
    archivePrefix = "arXiv",
    primaryClass = "gr-qc",
    doi = "10.1103/PhysRevD.94.124027",
    journal = "Phys. Rev. D",
    volume = "94",
    number = "12",
    pages = "124027",
    year = "2016"
}

@article{Ovalle:2023ref,
    author = "Ovalle, Jorge and Casadio, Roberto and Giusti, Andrea",
    title = "{Regular hairy black holes through Minkowski deformation}",
    eprint = "2304.03263",
    archivePrefix = "arXiv",
    primaryClass = "gr-qc",
    doi = "10.1016/j.physletb.2023.138085",
    journal = "Phys. Lett. B",
    volume = "844",
    pages = "138085",
    year = "2023"
}

@article{Mazza:2023iwv,
    author = "Mazza, Jacopo and Liberati, Stefano",
    title = "{Regular black holes and horizonless ultra-compact objects in Lorentz-violating gravity}",
    eprint = "2301.04697",
    archivePrefix = "arXiv",
    primaryClass = "gr-qc",
    doi = "10.1007/JHEP03(2023)199",
    journal = "JHEP",
    volume = "03",
    pages = "199",
    year = "2023"
}

@article{Modesto:2010rv,
    author = "Modesto, Leonardo and Nicolini, Piero",
    title = "{Charged rotating noncommutative black holes}",
    eprint = "1005.5605",
    archivePrefix = "arXiv",
    primaryClass = "gr-qc",
    doi = "10.1103/PhysRevD.82.104035",
    journal = "Phys. Rev. D",
    volume = "82",
    pages = "104035",
    year = "2010"
}

@article{Casadio:2023iqt,
    author = "Casadio, Roberto and Giusti, Andrea and Ovalle, Jorge",
    title = "{Quantum rotating black holes}",
    eprint = "2303.02713",
    archivePrefix = "arXiv",
    primaryClass = "gr-qc",
    doi = "10.1007/JHEP05(2023)118",
    journal = "JHEP",
    volume = "05",
    pages = "118",
    year = "2023"
}

@article{Lewandowski:2022zce,
    author = "Lewandowski, Jerzy and Ma, Yongge and Yang, Jinsong and Zhang, Cong",
    title = "{Quantum Oppenheimer-Snyder and Swiss Cheese Models}",
    eprint = "2210.02253",
    archivePrefix = "arXiv",
    primaryClass = "gr-qc",
    doi = "10.1103/PhysRevLett.130.101501",
    journal = "Phys. Rev. Lett.",
    volume = "130",
    number = "10",
    pages = "101501",
    year = "2023"
}

@article{Carballo-Rubio:2019fnb,
    author = "Carballo-Rubio, Ra{\'u}l and Di Filippo, Francesco and Liberati, Stefano and Visser, Matt",
    title = "{Geodesically complete black holes}",
    eprint = "1911.11200",
    archivePrefix = "arXiv",
    primaryClass = "gr-qc",
    doi = "10.1103/PhysRevD.101.084047",
    journal = "Phys. Rev. D",
    volume = "101",
    pages = "084047",
    year = "2020"
}

@article{Bonanno:2000ep,
    author = "Bonanno, Alfio and Reuter, Martin",
    title = "{Renormalization group improved black hole space-times}",
    eprint = "hep-th/0002196",
    archivePrefix = "arXiv",
    reportNumber = "INFN-CT-03-00, MZ-TH-00-04",
    doi = "10.1103/PhysRevD.62.043008",
    journal = "Phys. Rev. D",
    volume = "62",
    pages = "043008",
    year = "2000"
}

@article{Torres:2017ygl,
    author = "Torres, Ramon",
    title = "{Nonsingular black holes, the cosmological constant, and asymptotic safety}",
    eprint = "1703.09997",
    archivePrefix = "arXiv",
    primaryClass = "gr-qc",
    doi = "10.1103/PhysRevD.95.124004",
    journal = "Phys. Rev. D",
    volume = "95",
    number = "12",
    pages = "124004",
    year = "2017"
}

@article{Eichhorn:2012va,
    author = "Eichhorn, Astrid",
    title = "{Quantum-gravity-induced matter self-interactions in the asymptotic-safety scenario}",
    eprint = "1204.0965",
    archivePrefix = "arXiv",
    primaryClass = "gr-qc",
    doi = "10.1103/PhysRevD.86.105021",
    journal = "Phys. Rev. D",
    volume = "86",
    pages = "105021",
    year = "2012"
}

@article{Pawlowski:2023dda,
    author = {Pawlowski, Jan M. and Tr{\"a}nkle, Jan},
    title = "{Effective action and black hole solutions in asymptotically safe quantum gravity}",
    eprint = "2309.17043",
    archivePrefix = "arXiv",
    primaryClass = "hep-th",
    doi = "10.1103/PhysRevD.110.086011",
    journal = "Phys. Rev. D",
    volume = "110",
    number = "8",
    pages = "086011",
    year = "2024"
}

@article{Stashko:2024wuq,
    author = "Stashko, Oleksandr",
    title = "{Quasinormal modes and gray-body factors of regular black holes in asymptotically safe gravity}",
    eprint = "2407.07892",
    archivePrefix = "arXiv",
    primaryClass = "gr-qc",
    doi = "10.1103/PhysRevD.110.084016",
    journal = "Phys. Rev. D",
    volume = "110",
    number = "8",
    pages = "084016",
    year = "2024"
}

@article{Berti:2015itd,
    author = "Berti, Emanuele and others",
    title = "{Testing General Relativity with Present and Future Astrophysical Observations}",
    eprint = "1501.07274",
    archivePrefix = "arXiv",
    primaryClass = "gr-qc",
    doi = "10.1088/0264-9381/32/24/243001",
    journal = "Class. Quant. Grav.",
    volume = "32",
    pages = "243001",
    year = "2015"
}

@article{Yunes:2013dva,
    author = "Yunes, Nicol{\'a}s and Siemens, Xavier",
    title = "{Gravitational-Wave Tests of General Relativity with Ground-Based Detectors and Pulsar Timing-Arrays}",
    eprint = "1304.3473",
    archivePrefix = "arXiv",
    primaryClass = "gr-qc",
    doi = "10.12942/lrr-2013-9",
    journal = "Living Rev. Rel.",
    volume = "16",
    pages = "9",
    year = "2013"
}

@article{Markov:1985py,
    author = "Markov, M. A. and Mukhanov, Viatcheslav F.",
    title = "{DE SITTER LIKE INITIAL STATE OF THE UNIVERSE AS A RESULT OF ASYMPTOTIC DISAPPEARANCE OF GRAVITATIONAL INTERACTIONS OF MATTER}",
    doi = "10.1007/BF02732276",
    journal = "Nuovo Cim. B",
    volume = "86",
    pages = "97--102",
    year = "1985"
}

@article{Gao:2024cgg,
    author = "Gao, Xiao-Jun",
    title = "{Probing a regular black hole within asymptotically safe gravity via strong gravitational lensings and optical appearances}",
    eprint = "2411.09513",
    archivePrefix = "arXiv",
    primaryClass = "gr-qc",
    month = "11",
    year = "2024"
}

@inproceedings{Spina:2024npx,
    author = "Spina, Andrea and Silveravalle, Samuele and Bonanno, Alfio",
    title = "{Scalar Perturbations of Regular Black Holes derived from a Non-Singular Collapse Model in Asymptotic Safety}",
    booktitle = "17th Marcel Grossmann Meeting: {On Recent Developments in Theoretical and Experimental General Relativity, Gravitation, and Relativistic Field Theories}",
    eprint = "2410.05936",
    archivePrefix = "arXiv",
    primaryClass = "gr-qc",
    month = "10",
    year = "2024"
}

@article{Bhattacharjee:2025xcb,
    author = "Bhattacharjee, Chirantana and Sau, Subhadip and Mukherjee, Avijit",
    title = "{Radiative and jet signatures of regular black holes in quantum-corrected gravity}",
    eprint = "2509.26366",
    archivePrefix = "arXiv",
    primaryClass = "gr-qc",
    doi = "10.1140/epjc/s10052-025-14725-6",
    journal = "Eur. Phys. J. C",
    volume = "85",
    number = "9",
    pages = "1071",
    year = "2025"
}

@article{Mustafa:2025cou,
    author = "Mustafa, G. and Alimova, Asalkhon and Atamurotov, Farruh and Ibraheem, Awad A. and Channuie, Phongpichit and Bahaddinova, Gunel",
    title = "{Testing regular black holes in the framework of asymptotically safe gravity using particle dynamics, QPOs, and shadow constraints}",
    doi = "10.1140/epjc/s10052-025-14431-3",
    journal = "Eur. Phys. J. C",
    volume = "85",
    number = "7",
    pages = "741",
    year = "2025"
}

@article{Mannobova:2025uqf,
    author = "Mannobova, Sojida and Atamurotov, Farruh and Abdujabbarov, Ahmadjon and Alkahtani, Badr S. and Mustafa, G.",
    title = "{Spinning particle motion around asymptotically safe gravity exhibiting regular black holes}",
    doi = "10.1140/epjc/s10052-025-14305-8",
    journal = "Eur. Phys. J. C",
    volume = "85",
    number = "5",
    pages = "586",
    year = "2025"
}

@article{Urmanov:2025nou,
    author = "Urmanov, Abdybek and Chakrabarty, Hrishikesh and Malafarina, Daniele",
    title = "{Observational properties of regular black holes in asymptotic safety}",
    eprint = "2504.12072",
    archivePrefix = "arXiv",
    primaryClass = "gr-qc",
    doi = "10.1140/epjc/s10052-025-14377-6",
    journal = "Eur. Phys. J. C",
    volume = "85",
    number = "6",
    pages = "642",
    year = "2025"
}

@article{Zhao:2025sck,
    author = "Zhao, Lai and Tang, Meirong and Xu, Zhaoyi",
    title = "{Constraints on the scale parameter of regular black hole in asymptotically safe gravity from extreme mass ratio inspirals}",
    eprint = "2503.06503",
    archivePrefix = "arXiv",
    primaryClass = "gr-qc",
    doi = "10.1088/1475-7516/2025/10/002",
    journal = "JCAP",
    volume = "10",
    pages = "002",
    year = "2025"
}

@article{Turakhonov:2025ojy,
    author = "Turakhonov, Ziyodulla and Atamurotov, Farruh and Ghosh, Sushant G. and Abdujabbarov, Ahmadjon",
    title = "{Probing effects of plasma on shadow and weak gravitational lensing by regular black holes in asymptotically safe gravity}",
    doi = "10.1016/j.dark.2025.101880",
    journal = "Phys. Dark Univ.",
    volume = "48",
    pages = "101880",
    year = "2025"
}

\end{document}